\documentclass[12pt]{article}

\usepackage{amsmath,amsfonts,enumerate,array}
\usepackage{graphicx}
\usepackage[usenames]{color}
\usepackage[english]{babel}
\usepackage{fancybox}
\usepackage{chngpage} 
\usepackage{url}

\usepackage{cite}

\newcommand{\captionfonts}{\small}

\makeatletter  
\long\def\@makecaption#1#2{%
  \vskip\abovecaptionskip
  \sbox\@tempboxa{{\captionfonts #1: #2}}%
 \ifdim \wd\@tempboxa >\hsize
    {\captionfonts #1: #2\par}
  \else
    \hbox to\hsize{\hfil\box\@tempboxa\hfil}%
  \fi
  \vskip\belowcaptionskip}
\makeatother   

\usepackage{mciteplus}
\usepackage[colorlinks=true,      linkcolor=blue,      urlcolor=blue,      filecolor=blue,      citecolor=blue,
      pdfstartview=FitH,     pdfpagemode=UseNone,      bookmarksopen=true]{hyperref}  
      
\usepackage[all]{hypcap}     

\begin{document}

\numberwithin{equation}{section}


\mathchardef\mhyphen="2D

\newcommand{\be}{\begin{equation}} 
\newcommand{\ee}{\end{equation}} 
\newcommand{\bea}{\begin{eqnarray}\displaystyle}
\newcommand{\eea}{\end{eqnarray}}
\newcommand{\bt}{\begin{tabular}}
\newcommand{\et}{\end{tabular}}
\newcommand{\bs}{\begin{split}}
\newcommand{\es}{\end{split}}

\renewcommand{\a}{\alpha}	
\renewcommand{\b}{\beta}
\newcommand{\g}{\gamma}		
\newcommand{\G}{\Gamma}
\renewcommand{\d}{\delta}
\newcommand{\D}{\Delta}
\renewcommand{\c}{\chi}			
\newcommand{\C}{\Chi}
\newcommand{\p}{\psi}			
\renewcommand{\P}{\Psi}
\newcommand{\s}{\sigma}		
\renewcommand{\S}{\Sigma}
\renewcommand{\t}{\tau}		
\newcommand{\e}{\epsilon}
\newcommand{\n}{\nu}
\newcommand{\m}{\mu}
\renewcommand{\r}{\rho}
\renewcommand{\l}{\lambda}

\newcommand{\nn}{\nonumber\\} 		
\newcommand{\newotimes}{}  				
\newcommand{\diff}{\,\text{d}}		
\newcommand{\h}{{1\over2}}				
\newcommand{\Gf}[1]{\G \Big{(} #1 \Big{)}}	
\newcommand{\floor}[1]{\left\lfloor #1 \right\rfloor}
\newcommand{\ceil}[1]{\left\lceil #1 \right\rceil}

\def\cA{{\cal A}} \def\cB{{\cal B}} \def\cC{{\cal C}}
\def\cD{{\cal D}} \def\cE{{\cal E}} \def\cF{{\cal F}}
\def\cG{{\cal G}} \def\cH{{\cal H}} \def\cI{{\cal I}}
\def\cJ{{\cal J}} \def\cK{{\cal K}} \def\cL{{\cal L}}
\def\cM{{\cal M}} \def\cN{{\cal N}} \def\cO{{\cal O}}
\def\cP{{\cal P}} \def\cQ{{\cal Q}} \def\cR{{\cal R}}
\def\cS{{\cal S}} \def\cT{{\cal T}} \def\cU{{\cal U}}
\def\cV{{\cal V}} \def\cW{{\cal W}} \def\cX{{\cal X}}
\def\cY{{\cal Y}} \def\cZ{{\cal Z}}

\def\mC{\mathbb{C}} \def\mP{\mathbb{P}}  
\def\mR{\mathbb{R}} \def\mZ{\mathbb{Z}} 
\def\mT{\mathbb{T}} \def\mN{\mathbb{N}}
\def\mH{\mathbb{H}} \def\mX{\mathbb{X}}
\def\CP{\mathbb{CP}}
\def\RP{\mathbb{RP}}
\def\Z{\mathbb{Z}}
\def\N{\mathbb{N}}
\def\H{\mathbb{H}}

\newcommand{\Zd}{\ensuremath{ Z^{\dagger}}}
\newcommand{\Xd}{\ensuremath{ X^{\dagger}}}
\newcommand{\Ad}{\ensuremath{ A^{\dagger}}}
\newcommand{\Bd}{\ensuremath{ B^{\dagger}}}
\newcommand{\Ud}{\ensuremath{ U^{\dagger}}}
\newcommand{\Td}{\ensuremath{ T^{\dagger}}}
\newcommand{\T}[3]{\ensuremath{ #1{}^{#2}_{\phantom{#2} \! #3}}}		
\newcommand{\tr}{\operatorname{tr}}
\newcommand{\sech}{\operatorname{sech}}
\newcommand{\Spin}{\operatorname{Spin}}
\newcommand{\Sym}{\operatorname{Sym}}
\newcommand{\Com}{\operatorname{Com}}
\def\adj{\textrm{adj}}
\def\id{\textrm{id}}
\def\pb{\ov\psi}
\def\pt{\widetilde{\psi}}
\def\at{\widetilde{\a}}
\def\cb{\ov\chi}
\def\db{\bar\partial}
\def\delb{\bar\partial}
\def\dbar{\ov\partial}
\def\dag{\dagger}
\def\dalpha{{\dot\alpha}}
\def\dbeta{{\dot\beta}}
\def\dgamma{{\dot\gamma}}
\def\ddelta{{\dot\delta}}
\def\ad{{\dot\alpha}}
\def\bd{{\dot\beta}}
\def\dg{{\dot\gamma}}
\def\dd{{\dot\delta}}
\def\th{\theta}
\def\Th{\Theta}
\def\eb{{\ov \epsilon}}
\def\gb{{\ov \gamma}}
\def\wb{{\ov w}}
\def\Wb{{\ov W}}
\def\D{\Delta}
\def\DD{\Delta^\dag}
\def\Db{\ov D}
\def\ov{\overline}
\def\Slash{\, / \! \! \! \!}
\def\dslash{\partial\!\!\!/} 
\def\Dslash{D\!\!\!\!/\,\,}
\def\fslash#1{\slash\!\!\!#1}
\def\Fslash#1{\slash\!\!\!\!#1}
\def\del{\partial}
\def\delb{\bar\partial}
\newcommand{\ex}[1]{{\rm e}^{#1}} 
\def\ii{{i}}
\newcommand{\vs}[1]{\vspace{#1 mm}}
\newcommand{\ve}{{\vec{\e}}}
\newcommand{\shalf}{\frac{1}{2}}
\newcommand{\lb}{\rangle}
\newcommand{\al}{\ensuremath{\alpha'}}
\newcommand{\ap}{\ensuremath{\alpha'}}
\newcommand{\ft}[2]{{\textstyle {\frac{#1}{#2}} }}

\newcommand{\rmd}{\mathrm{d}}
\newcommand{\rmx}{\mathrm{x}}
\def\tA{ {\widetilde A} } 
\def\one{{\hbox{\kern+.5mm 1\kern-.8mm l}}}
\def\zero{{\hbox{0\kern-1.5mm 0}}}
\def\eq#1{(\ref{#1})}
\newcommand{\secn}[1]{Section~\ref{#1}}
\newcommand{\tbl}[1]{Table~\ref{#1}}
\newcommand{\fig}{Fig.~\ref}
\def\sqi{{1\over \sqrt{2}}}
\newcommand{\hsp}{\hspace{0.5cm}}
\def\half{{\textstyle{1\over2}}}
\let\ci=\cite \let\re=\ref
\let\se=\section \let\sse=\subsection \let\ssse=\subsubsection
\newcommand{\dpb}{D$p$-brane}
\newcommand{\dpbs}{D$p$-branes}
\def\gh{{\rm gh}}
\def\sgh{{\rm sgh}}
\def\NS{{\rm NS}}
\def\R{{\rm R}}
\def\Qp{Q_{\rm P}}
\def\QP{Q_{\rm P}}
\newcommand\dott[2]{#1 \! \cdot \! #2}
\def\eo{\overline{e}}
\newcommand{\bb}{\bigskip}
\newcommand{\ac}[2]{\ensuremath{\{ #1, #2 \}}}
\renewcommand{\ell}{l}
\newcommand{\z}{\ell}
\newcommand{\bm}{\bibitem}

\def\b{\bigskip}
\def\I{\text{I}}
\def\II{\text{II}}

\begin{flushright}
\end{flushright}
\vspace{20mm}
\begin{center}
{\LARGE Partial Spectral Flow in the D1D5 CFT}
\\
\vspace{18mm}
\textbf{Bin} \textbf{Guo}{\footnote{bin.guo@ipht.fr}}~\textbf{and} ~ \textbf{Shaun}~  \textbf{Hampton}{\footnote{shaun.hampton@ipht.fr}}
\\
\vspace{10mm}

${}$Institut de Physique Th\'eorique,\\
	Universit\'e Paris-Saclay,
	CNRS, CEA, \\ 	Orme des Merisiers,\\ Gif-sur-Yvette, 91191 CEDEX, France  \\

\vspace{8mm}
\end{center}

\vspace{4mm}

\thispagestyle{empty}
\begin{abstract}
The two-dimensional $\mathcal{N}=4$ superconformal algebra has a free field realization with four bosons and four fermions. There is an automorphism of the algebra called spectral flow.
Under spectral flow, the four fermions are transformed together. 
In this paper, we study {\it partial spectral flow} where only two of the four fermions are transformed. 
Partial spectral flow is applied to the D1D5 CFT where a marginal deformation moves the CFT away from the free point. 
The partial spectral flow is broken by the deformation. 
We show that this effect can be studied due to a transformation of the deformation which is well-defined under partial spectral flow.
As a result in the spectrum, we demonstrate how to compute the second-order energy lift of D1D5P states from their partial spectral flowed states. 
We find that D1D5P states related by partial spectral flow do not have the same lift in general.

\vspace{3mm}

\end{abstract}
\newpage

\setcounter{page}{1}

\numberwithin{equation}{section} 

\tableofcontents

\newpage

\section{Introduction}
The D1D5 CFT is a $1+1$ dimensional theory with $\mathcal{N} = (4,4)$ superconformal symmetry. Using the AdS/CFT correspondence \cite{Maldacena:1997re,Gubser:1998bc,Witten:1998qj}, it has proven extremely useful in describing many aspects of black hole physics \cite{Callan:1996dv,Das:1996wn,Das:1996ug,Maldacena:1996ix,David:1999ec}. 
There is a particular location in the moduli space of parameters called the `orbifold point' in which the CFT is free \cite{Seiberg:1999xz,Dijkgraaf:1998gf,Larsen:1999uk,Jevicki:1998bm}. At this point many interesting properties of black holes have been obtained. It has given a field theory derivation of the Bekenstein Hawking entropy \cite{Strominger:1996sh,Maldacena:1999bp}. The spectrum of low energy Hawking radiation was obtained in \cite{Das:1996wn,Maldacena:1996ix}. Furthermore, the orbifold point has been used to obtain properties of black hole microstate geometries \cite{Lunin:2001fv,Lunin:2001jy,Mathur:2005zp,Bena:2015bea,Bena:2016agb,Bena:2016ypk,Bena:2017xbt,Ceplak:2018pws,Heidmann:2019zws,Kanitscheider:2006zf,Kanitscheider:2007wq,Taylor:2007hs,Giusto:2015dfa,GarciaTormo:2019inl,Giusto:2019qig,Rawash:2021pik,Ganchev:2021ewa}. More recently progress has been made in deriving the tensionless string limit of the AdS/CFT correspondence using the orbifold theory \cite{Eberhardt:2018ouy,Eberhardt:2019ywk,Eberhardt:2020akk,Dei:2020zui,Knighton:2020kuh,Gaberdiel:2021kkp,Eberhardt:2021vsx}. 

The left (right) moving part of the $\mathcal{N} = (4,4)$ superconformal algebra (SCA) is 
the $\mathcal{N} = 4$ SCA \cite{Schwimmer:1986mf,Sevrin:1988ew}. It has a free field realization composed of four bosons and four fermions. The basic commutators of the fermionic modes are
\be\label{b com intro}
\lbrace d^{++}_r , d^{--}_s\rbrace  =- \delta_{r+s,0},~~~~~~~~~\lbrace d^{+-}_r , d^{-+}_s\rbrace  =\delta_{r+s,0}
\ee
A very important property of the $\mathcal{N} = 4$ SCA, which we shall investigate in this paper, is an automorphism called spectral flow \cite{Schwimmer:1986mf}. It acts on the elements of an algebra transforming them in a way that keeps the algebra itself invariant.
For some applications in microstate geometries, see \cite{Jejjala:2005yu,Giusto:2012yz,Chakrabarty:2015foa}.
In the free field realization, it can be simply understood as the following transformation
\be\label{flow intro}
d^{\pm A}_r\to d^{\pm A}_{r\mp{\alpha\over2}}
\ee
where $A=+,-$ and $\alpha$ is a real parameter. Bosonic fields are unaffected by this transformation. 
Spectral flow (\ref{flow intro}) keeps the basic commutators (\ref{b com intro}) invariant and thus keeps the $\mathcal{N} = 4$ SCA invariant.
When $\alpha$ is an odd integer, the spectral flow interpolates between the Neveu-Schwarz (NS) sector and the Ramond (R) sector of the fermionic degrees of freedom. 

Within spectral flow, the four fermions transform together. However, in the free field realization (\ref{b com intro}) two of the four fermions are decoupled from the other two. Thus we can perform spectral flow for only two of the four fermions while still keeping the basic commutators (\ref{b com intro})
invariant therefore keeping the $\mathcal{N} = 4$ SCA invariant. We will call it {\it partial spectral flow}.
In this paper, we will study the structure of this partial spectral flow. We will show that it is related to the existence of two $\mathcal{N} = 2$ SCA's in the free field realization.

The orbifold (free) point of the D1D5 CFT can be obtained by orbifolding $N=n_1n_5$ copies of the above free field realization with the permutation group $S_{N}$. In the orbifold theory, there exists a particular interesting class of operators called twist operators. A twist operator $\sigma_k$ of rank $k$ can link $k$ copies of the free fields into a single copy of winding $k$. We will study the transformation of twist operators under partial spectral flow. We will show that they have well-defined transformations.

To move towards the supergravity point one must deform away from the orbifold point using a marginal deformation. See \cite{David:1999ec,Gomis:2002qi,Gava:2002xb,Avery:2010qw}
for more details. 
We will show that the marginal deformation, which contains the twist operator $\sigma_2$ of rank 2, is not invariant under partial spectral flow. Thus partial spectral flow is broken explicitly by the marginal deformation. We study the effect of this breaking on the spectrum of D1D5P states which are states with only left moving modes excited. We show that states related by partial spectral flow do not have the same energy lift in general. We will give an explicit example. This example also shows that partial spectral flow can be used to simplify lifting computations.

The outline of the paper is as follows. 
In section \ref{sec SCA} we write down the two-dimensional $\mathcal{N}=4$ SCA, its free field realization and spectral flow. We will use a formalism that includes the orbifold theory naturally. In section \ref{sec partial} 
we introduce partial spectral flow and analyze its structure. 
In section \ref{sec deformation} we study the transformation of twist operators under partial spectral flow. We will also show that partial spectral flow is broken by the marginal deformation.
In section \ref{lifting} we study the effect of the breaking of partial spectral flow on the spectrum of D1D5P states.
In section \ref{conclusion} we conclude.

\section{Two-dimensional $\mathcal{N}=4$ superconformal algebra}\label{sec SCA}

In this section we recall the two-dimensional $\mathcal{N}=4$ SCA and its realization by using four free bosons and four free fermions.  In the following, we will use the formalism (see for example \cite{Hampton:2018ygz,Guo:2020gxm}) that also includes the $k$-twisted sector of the orbifold theory. For readers who are only interested in partial spectral flow but not the orbifold theory, set $k=1$.

\subsection{Commutation relations}
 
The $\mathcal{N}=4$ SCA contains the Virasoro generators $L_m$ which form a Virasoro algebra and $SU(2)$ current generators $J^a_m$ with $a=1,2,3$ which form an $SU(2)$ Kac-Moody algebra. There are also four supercharge generators $G_{\dot A,r}^{\a}$ with $\dot A =+,-$ and $\a = +,-$,
\footnote{This $\alpha$ should not be confused with the spectral flow parameter $\alpha$.}
which form two $SU(2)$ doublets with charges $\alpha/2$. 

The commutation relations are
\bea\label{com}
[L_m,L_n] &=& {c\over12}m(m^2-1)\delta_{m+n,0}+ (m-n)L_{m+n}\cr
[J^a_{m},J^b_{n}] &=&{c\over12}m\delta^{ab}\delta_{m+n,0} +  i\e^{ab}_{\,\,\,\,c}J^c_{m+n}\cr
[L_{m},J^a_n]&=& -nJ^a_{m+n}\cr
[L_{m},G^{\a}_{\dot{A},r}] &=& ({m\over2}  -r)G^{\a}_{\dot{A},m+r}\cr
[J^a_{m},G^{\a}_{\dot{A},r}] &=&{1\over2}(\s^{aT})^{\a}_{\beta} G^{\beta}_{\dot{A},m+r}\cr
\lbrace G^{\a}_{\dot{A},r} , G^{\beta}_{\dot{B},s} \rbrace&=&  \e_{\dot{A}\dot{B}}\bigg[\e^{\a\beta}{c\over6}(r^2-{1\over4})\delta_{r+s,0}  + (\s^{aT})^{\a}_{\g}\e^{\g\beta}(r-s)J^a_{r+s}  + \e^{\a\beta}L_{r+s}  \bigg]
\eea
where $c=6k$. Here $k$ is a positive integer and is also the winding number of the $k$-twisted sector in the orbifold theory that will be introduced later. Let $q$ be an integer. The mode numbers for the bosonic generators $L$ and $J$ are $n=q/k$. The mode numbers for the fermionic generators $G$ are $r=q/k$ in the R sector and $r=(q+\frac{1}{2})/k$ in the NS sector.

It is convenient to use $J^\pm_n$ defined as
\bea
J^\pm_n=J^1_n \pm i J^2_n
\eea
Then we have
\bea
[J^3_m , J^{+}_n] &=& J^{+}_{m+n},\qquad\qquad [J^3_m , J^{-}_n] ~=~ -J^{-}_{m+n}\cr
[J^+_{m},J^-_{n}]&=&{c\over6}m\delta_{m+n,0} + 2J^3_{m+n}\cr
[J^{3}_{m},G^{\pm}_{\dot{A},r}]  &=& \pm \frac{1}{2}G^{\pm}_{\dot{A},m+r} \cr
[J^{+}_{m},G^{+}_{\dot{A},r}]  &=& 0 ,\qquad\qquad ~~~[J^{-}_{m},G^{+}_{\dot{A},r}]  ~=~ G^{-}_{\dot{A},m+r}\cr
[J^{+}_{m},G^{-}_{\dot{A},r}]  &=&G^{+}_{\dot{A},m+r},\qquad ~[J^{-}_{m},G^{-}_{\dot{A},r}]  ~=~ 0 
\eea

\subsection{The free field realization}

The $\mathcal{N}=4$ superconformal algebra (\ref{com}) can be realized by using four free bosons 
$\alpha_{A\dot A,n}$ with $A,\dot A=+,-$ and four free fermions $d^{\alpha A}_r$ with $\alpha, A=+,-$. The basic commutation relations are
\bea\label{basic com}
[\a_{A\dot{A},m},\a_{B\dot{B},n}] &=& - k m\e_{A\dot{A}}\e_{B\dot{B}}\delta_{m+n,0}\cr
\lbrace d^{\alpha A}_r , d^{\beta B}_s\rbrace  &=&- k \e^{\alpha\beta}\e^{AB}\delta_{r+s,0}
\eea
The mode numbers for the bosons $\alpha$ are $n=q/k$. The mode numbers for the fermions $d$ are $r=q/k$ in the R sector and $r=(q+\frac{1}{2})/k$ in the NS sector. Notice that since the spacing of mode numbers is $1/k$ there are effectively $4k$ bosons and $4k$ fermions with total central charge $6k$ which is equal to the central charge $c$ in the algebra (\ref{com}).

The generators of the $\mathcal{N}=4$ SCA can be realized as
\bea\label{g}
J^a_m &=& {1\over 4 k}\sum_{r}\epsilon_{AB}d^ {\g B}_r\epsilon_{\alpha\gamma}(\s^{aT})^{\a}_{\beta}d^ {\beta A}_{m-r},\qquad a=1,2,3\cr
J^3_m &=&  - {1\over 2 k}\sum_{r} d^ {+ +}_{r}d^ {- -}_{m-r} - {1\over 2 k}\sum_{r}d^ {- +}_r d^ {+ -}_{m-r}\cr
J^{+}_m&=& {1\over k}\sum_{r}d^ {+ +}_rd^ {+ -}_{m-r} ,\qquad J^{-}_m= {1\over k} \sum_{r}d^ {--}_rd^ {- +}_{m-r}\cr
G^{\a}_{\dot{A},r} &=& - {i\over k} \sum_{n}d^ {\a A}_{r-n} \a_{A\dot{A},n}\cr
L_m&=&\! -{1\over 2 k}\sum_{n} \e^{AB}\e^{\dot A \dot B}\a_{A\dot{A},n}\a_{B\dot{B},m-n}- {1\over 2 k}\sum_{r}(m-r+{1\over2})\epsilon_{\alpha\beta}\epsilon_{AB}d^ {\a A}_r d^ {\beta B}_{m-r}
\eea
There is an implicit normal ordering in the above definition.
By using the basic commutators (\ref{basic com}) one can check that the above generators satisfy the $\mathcal{N}=4$ SCA (\ref{com}). 
We can also find the commutation relations between the free fields and the $\mathcal{N}=4$ superconformal generators, which are
\bea\label{commutations}
[L_m,\a_{A\dot{A},n}] &=&-n\a_{A\dot{A},m+n} \cr
[L_m ,d^{\a A}_r] &=&-({m\over2}+r)d^{\a A}_{m+r}\cr
\lbrace G^{\a}_{\dot{A},r} ,  d^{\beta B}_{s} \rbrace&=&i\e^{\a\beta}\e^{AB}\a_{A\dot{A},r+s}\cr
[G^{\a}_{\dot{A},r} , \a_{B \dot{B},m}]&=&  -im\e_{AB}\e_{\dot{A}\dot{B}}d^{\a A}_{r+m}\cr
[J^a_m,d^{\a A}_r] &=&{1\over 2}(\s^{Ta})^{\a}_{\beta}d^{\beta A}_{m+r}
\eea
In terms of $J^{\pm}$ and $J^3$, the last one becomes
\bea
[J^{3}_m,d^{\pm A}_r] &=& \pm \frac{1}{2} d^{\pm A}_{m+r}\cr
[J^{+}_m,d^{+ A}_r] &=& 0,\qquad~~~~~ [J^{-}_m,d^{+ A}_r] ~=~ d^{-A}_{m+r}\cr
[J^{-}_m,d^{+ A}_r] &=& d^{-A}_{m+r},\qquad [J^{+}_m,d^{+ A}_r] ~=~ 0
\eea
One can also normalize the basic commutation relations (\ref{basic com}) to remove the parameter $k$.
In this way, there is no overall $1/k$ factor in the generators (\ref{g}). All other commutation relations will be the same.

\subsection{Spectral flow}

Spectral flow is an automorphism of the $\mathcal{N}=4$ SCA. It is an operation which changes the boundary conditions of the fermionic fields within the theory, interpolating between the NS and R sectors \cite{Schwimmer:1986mf}. Under a spectral flow by $\a$ units, the transformations of the generators are
\bea\label{flow}
L_n&\to& L_n - \a J^{3}_n + {\a^2c\over 24}\delta_{n,0}\cr
J^{3}_n&\to& J^{3}_n-{\a c\over12}\delta_{n,0}\cr
G^{\pm}_{\dot A,r}&\to& G^{\pm}_{\dot A,r\mp{\a\over2}}\cr
J^{\pm}_n&\to& J^{\pm}_{n\mp{\alpha}}
\eea
It can be checked that the above spectral flow preserves the algebra (\ref{com}). This is why it is called an automorphism.

In the free field realization, the above spectral flow can be derived from the basic spectral flow of the free fermions
\bea\label{b flow}
d^{\pm A}_r&\to&d^{\pm A}_{r\mp{\alpha\over2}}
\eea
which preserves the basic commutation relations (\ref{basic com}).
Notice that some terms in the generators (\ref{g}) will be reordered under the basic spectral flow (\ref{b flow}) due to the normal ordering. This gives the terms containing the central charge in (\ref{flow}). 

Another way to see that the $\mathcal{N}=4$ SCA (\ref{com}) is invariant under spectral flow (\ref{flow}) is to use the free field realization. The $\mathcal{N}=4$ SCA can be derived from the basic commutation relations (\ref{basic com}). The basic spectral flow which preserves the basic commutation relations should also preserve the $\mathcal{N}=4$ SCA. Thus spectral flow (\ref{flow}) which comes from the basic spectral flow also preserves the $\mathcal{N}=4$ SCA. For an operator $O(z)$ of charge $j$, the transformation under spectral flow by $\alpha$ units at $z=0$ is
\be
O(z)\to z^{-\alpha j}O(z) 
\ee
In the following, we will also use the cylinder coordinate $w$ defined through $z=e^w$. Spectral flow at $w=-\infty$ becomes
\be
O(w)\to  e^{-\alpha j w}O(w)
\ee

\section{Partial spectral flow}\label{sec partial}

The fermionic part of the basic commutators (\ref{basic com}) is
\bea
\lbrace d^{+ +}_r , d^{- -}_s\rbrace  &=&- k \delta_{r+s,0}\nn
\lbrace d^{+ -}_r , d^{- +}_s\rbrace  &=& k \delta_{r+s,0}
\eea
The four fermions can be separated into two types with charges $++,- -$ and $+-,-+$. The two types decouple from each other. Thus instead of spectral flowing all the fermions together as in (\ref{b flow}), we can spectral flow the two types separately. The transformation is
\bea\label{flow 1}
d^{++}_r&\to&d^{++}_{r-{\alpha_{\I}\over2}}\cr
d^{--}_r&\to&d^{--}_{r+{\alpha_{\I}\over2}}
\eea
and
\bea\label{flow 2}
d^{+-}_r&\to&d^{+-}_{r-{\alpha_{\II}\over2}}\cr
d^{-+}_r&\to&d^{-+}_{r+{\alpha_{\II}\over2}}
\eea
where $\alpha_{\I}$ and $\alpha_{\II}$ are two independent free parameters. We will call this a \textit{partial spectral flow}. The purpose of this paper is to investigate its structure and its application in the D1D5 CFT.

\subsection{Decomposition of the generators}\label{reduction to subalgebra}

In this section we will decompose the generators in the $\mathcal{N}=4$ SCA (\ref{com}). 
Let us group the four bosons and four fermions into type I and type II as follows
\bea\label{type}
\text{Type I} : ~d^{++}_r~~~d^{--}_r~~~\a_{++,n}~~~\a_{--,n}\nn
\text{Type II} : ~d^{+-}_r~~~d^{-+}_r~~~\a_{+-,n}~~~\a_{-+,n}
\eea
Based on this grouping, let us decompose the generators in the free field realization of the $\mathcal{N}=4$ SCA (\ref{g}).

Let us begin with generators that can be decomposed. They are
\be\label{de}
L_m~~~~J^3_m~~~~G^{+}_{+,r}~~~G^{-}_{-,r}
\ee
For $L_m$ we have
\bea
L_m
&\equiv& L ^{(\I)}_m + L^{(\II)}_m
\eea
where
\bea
L^{(\I)}_{m}&=&-{1\over2k}\sum_r(m-r+{1\over2})\big(d^ {++}_r d^ {--}_{m-r} + d^ {--}_r d^ {++}_{m-r}\big)\cr
&&-{1\over2k}\sum_n\big(\a_{++,n}\a_{--,m-n} + \a_{--,n}\a_{++,m-n}  \big)\cr
L^{(\II)}_m&=&{1\over2k}\sum_r(m-r+{1\over2})\big(d^ {+-}_r d^ {-+}_{m-r} + d^ {-+}_r d^ {+-}_{m-r}\big)\cr
&&\hspace{-0.3cm}+{1\over2k}\sum_n\big(\a_{+-,n}\a_{-+,m-n} + \a_{-+,n}\a_{+-,m-n}  \big)
\eea
For $J^3_m$ we have
\bea
J^3_m &=&   - {1\over 2k}\sum_{r} d^ {+ +}_{r}d^ {- -}_{m-r} - {1\over 2k}\sum_{r}d^ {- +}_r d^ {+ -}_{m-r}\cr
&\equiv&J^{3(\I)}_m + J^{3(\II)}_m
\eea
For $G^{+}_{+,r}$ and $G^{-}_{-,r}$ we have
\bea\label{G++}
G^{+}_{+,r} 
&=&-{i\over k}\sum_{n}d^ {+ +}_{r-n} \a_{++,n}-i\sum_{n}d^ {+ -}_{r-n} \a_{-+,n}\cr
&\equiv&G^{+(\I)}_{+,r} + G^{+(\II)}_{+,r}
\cr
\cr
G^{-}_{-,r} 
&=&-{i\over k}\sum_{n}d^ {- -}_{r-n} \a_{--,n}-i\sum_{n}d^ {- +}_{r-n} \a_{+-,n}\cr
&\equiv&G^{-(\I)}_{-,r} + G^{-(\II)}_{-,r}
\eea

There are generators that cannot be decomposed into type I and type II because they are composed of one member from each of the two types respectively. They are
\be\label{nde}
J^{+}_m~~~~J^{-}_m~~~~G^{+}_{-,r}~~~~G^{-}_{+,r}
\ee
In detail, they are
\be
J^{+}_m={1\over k}\sum_{r}d^ {+ +}_rd^ {+ -}_{m-r} ~~~\qquad J^{-}_m={1\over k}\sum_{r}d^ {--}_rd^ {- +}_{m-r}
\ee
and
\bea\label{G+-}
G^{+}_{-,r} 
&=&-{i\over  k}\sum_{n}d^ {+ +}_{r-n} \a_{+-,n}-{i\over k}\sum_{n}d^ {+ -}_{r-n} \a_{--,n}\cr
&\equiv&G^{+(\I)}_{-,r} + G^{+(\II)}_{-,r}
\cr
\cr
G^{-}_{+,r} 
&=&-{i\over k}\sum_{n}d^ {- -}_{r-n} \a_{-+,n}-{i\over k}\sum_{n}d^ {- +}_{r-n} \a_{++,n}\cr
&\equiv&G^{-(\I)}_{+,r} + G^{-(\II)}_{+,r}
\eea
where we have assigned type I and type II for the terms in $G^{+}_{-,r}$ and $G^{-}_{+,r}$ based only on the fermions in (\ref{type}) for later convenience even though they cannot be decomposed.

Consider the generators that can be decomposed given in (\ref{de}).
We note that the modes $L^{(\I)},J^{3(\I)},G^{+(\I)}_{+},G^{-(\I)}_{-}$ close within themselves and form a $\mathcal{N}=2$ SCA with a $U(1)$ group and a central charge $c/2$. Similarly, $L^{(\II)},J^{3(\II)},G^{+(\II)}_{+},G^{-(\II)}_{-}$ form another $\mathcal{N}=2$ SCA with another $U(1)$ group and a central charge $c/2$.
They satisfy the following two $\mathcal{N}=2$ SCA's 
\bea\label{commutations_ii}
[L^{(i)}_m,L^{(j)}_n] &=&\d^{ij} \bigg[{c\over24}m(m^2-1)\delta_{m+n,0}+ (m-n)L^{(j)}_{m+n}\bigg]\cr
[J^{3(i)}_{m},J^{3(j)}_{n}] &=&\d^{ij}{c\over24}m\delta_{m+n,0} \cr
\lbrace G^{+(i)}_{+,r} , G^{-(j)}_{-,s} \rbrace&=&  -\d^{ij}\bigg[{c\over12}(r^2-{1\over4})\delta_{r+s,0}  + (r-s)J^{3(j)}_{r+s}  + L^{(j)}_{r+s}  \bigg]\cr
[J^{3(i)}_{m},G^{+(j)}_{+,r}] &=&\d^{ij}{1\over2} G^{+(j)}_{+,m+r}~~~~~~~~~~~~~
[J^{3(i)}_{m},G^{-(j)}_{-,r}] =-\d^{ij}{1\over2} G^{-(j)}_{-,m+r}\cr
[L^{(i)}_{m},J^{3(j)}_n]&=& -\d^{ij}nJ^{3(j)}_{m+n}\cr
[L^{(i)}_{m},G^{+(j)}_{+,r}] &=& \d^{ij}({m\over2}  -r)G^{+(j)}_{+,m+r}~~~~~
[L^{(i)}_{m},G^{-(j)}_{-,r}] = \d^{ij}({m\over2}  -r)G^{-(j)}_{-,m+r}
\eea
where $i,j=\I,\II$. Notice that these two $\mathcal{N}=2$ SCA's exist only in the free field realization where we can decompose the generators. The two partial spectral flows (\ref{flow 1}) and (\ref{flow 2}) of the $\mathcal{N}=4$ SCA are indeed the spectral flow for the two $\mathcal{N}=2$ SCA's.

In this section we decomposed the generators based on the grouping (\ref{type}). There is another way to group the bosonic fields. We can group  $d^{++}_r,d^{--}_r,\a_{+-,n},\a_{-+,n}$ into type I and $d^{+-}_r,d^{-+}_r,\a_{++,n},\a_{--,n}$ into type II. In this way, similar results will follow. The two $\mathcal{N}=2$ SCA's are formed by $L^{(i)},J^{3(i)},G^{+(i)}_{-},G^{-(i)}_{+}$ with $i=\I,\II$.

\subsection{Partial spectral flow}\label{partial spectral flow}

Using the free field realization, one can derive the transformation of the generators under partial spectral flow. We list the results in the following:
\subsubsection*{Partial spectral flow by $\a_\I$}
Partial spectral flow of type I for the free fields are
\bea
d^{++}_r&\to&d^{++}_{r-{\alpha_{\I}\over2}}\cr
d^{--}_r&\to&d^{--}_{r+{\alpha_{\I}\over2}}
\eea
It gives
\bea\label{flow 1 g}
L^{(\I)}_n&\to& L^{(\I)}_n - \a_\I J^{3(\I)}_n + {\a^2_\I c\over48}\delta_{n,0}\cr
J^{3(\I)}_n&\to& J^{3(\I)}_n-{\a_\I c\over24}\delta_{n,0}\cr
G^{\pm(\I)}_{\dot A,r}&\to& G^{\pm(\I)}_{\dot A,r\mp{\a_\I\over2}}\cr
J^{\pm}_n&\to& J^{\pm}_{n\mp{\alpha_{\I}\over2}}
\eea
with all other generators labelled by type II remaining unchanged. 
In the generator $J^{+}_n={1\over k}\sum_{r}d^ {+ +}_rd^ {+ -}_{n-r}$, we only spectral flow $d^{++}_r$. Then $J^{\pm}_n$ partial spectral flows to $J^{\pm}_{n\mp{\alpha_{\I}\over 2}}$ not $J^{\pm}_{n\mp{\alpha_{\I}}}$ as in spectral flow (\ref{flow}). Another way to understand this fact is the following. The $J^{\pm}_n$ has one unit of $\pm$ charge where a $1/2$ unit comes from type I and the other $1/2$ unit comes from type II. If we only partial spectral flow type I by $\alpha_{\I}$, the mode number of  $J^{\pm}_n$ is shifted by $\alpha_{\I}/2$ not $\alpha_{\I}$. Similar ideas will be used later to derive the transformation of twist operators in the orbifold theory. The transformations for the quantum numbers can be derived
\bea\label{1 qn}
{h^{(\I)}}'&=& h^{(\I)} + \a_\I j^{(\I)} + {\a_\I^{2} c\over 48}\cr
{j^{(\I)}}' &=& j^{(\I)} + {\a_\I c\over 24}
\eea

\subsubsection*{Partial spectral flow by $\a_\II$}
Partial spectral flow of type II for the free fields are
\bea
d^{+-}_r&\to&d^{+-}_{r-{\alpha_{\II}\over2}}\cr
d^{-+}_r&\to&d^{-+}_{r+{\alpha_{\II}\over2}}
\eea
It gives
\bea\label{flow 2 g}
L^{(\II)}_n&\to& L^{(\II)}_n - \a_\II J^{3(\II)}_n  + {\a_\II^2c\over48}\delta_{n,0}\cr
J^{3(\II)}_n&\to& J^{3(\II)}_n-{\a_\II c\over24}\delta_{n,0}\cr
G^{\pm(\II)}_{\dot A,r}&\to& G^{\pm(\II)}_{\dot A,r\mp{\a_\II\over2}}\cr
J^{\pm}_n&\to& J^{\pm}_{n\mp{\alpha_{\II}\over2}}
\eea
with all other operators labelled by type I remaining unchanged. 
The transformations for the quantum numbers are
\bea\label{2 qn}
{h^{(\II)}}'&=& h^{(\II)} + \a_\II j^{(\II)} + {\a_\II^{2} c\over 48}\cr
{j^{(\II)}}' &=& j^{(\II)} + {\a_\II c\over 24}
\eea
Notice that the transformation of quantum numbers (\ref{1 qn}) and (\ref{2 qn}) are indeed the ones for the spectral flow of the two $\mathcal{N}=2$ SCA's with central charge $c/2$.

As explained, the transformation rules (\ref{flow 1 g}) and (\ref{flow 2 g}) should preserve the $\mathcal{N}=4$ SCA since they preserve the basic commutation relations (\ref{basic com}). In the appendix, we give some explicit examples.

\section{Breaking of partial spectral flow}\label{sec deformation}

\subsection{The D1D5 CFT and the twist operator}
Consider Type IIB string theory on the space $M_{4,1}\times S^1\times T^4$ with $n_1$ D1 branes wrapping $S^1$ and $n_5$ $D5$ branes wrapping $S^1\times T^4$. In the near horizon limit the spacetime becomes $AdS_3\times S^3\times T^4$ where the $S^1$ is part of $AdS_3$. The D1D5 CFT is defined on the boundary of this $AdS_3$ region in the decoupling limit.

There is a point in the moduli space of couplings called the `orbifold point'. To build the theory at this point, let us first make a tensor product of $N=n_1n_5$ copies of the above free field realization of the $\mathcal{N}=4$ SCA with $k=1$. Then we take the orbifold by the permutation group $S_N$, which permutes the $N$ copies. In this $S_N$ orbifold theory, there exist $k$-twisted sectors where $k$ copies are joined together. 
They can be viewed as `component strings' with winding $k$.
The free bosons and fermions living in these $k$-twisted sectors have basic commutators (\ref{basic com}) with parameter $k$ and mode spacing $1/k$. 

There exist the following chiral primaries in these $k$-twisted sectors. For more details about their construction, see \cite{Lunin:2000yv,Lunin:2001pw,Pakman:2009zz,Pakman:2009ab,Pakman:2009mi}. For related works involving the twist operator, see \cite{Burrington:2012yq,Burrington:2012yn,Carson:2017byr,Burrington:2017jhh,Burrington:2018upk,Keller:2019yrr,Dei:2019iym}.
Their dimensions and charges are
\bea\label{chiral}
\sigma_{k}^{-}:&& h = j = \frac{k-1}{2},~~~j^{(\I)}= \frac{k-1}{4},~~~j^{(\II)}= \frac{k-1}{4}\nn
d^{++}_{-\frac{1}{2}}\sigma_{k}^{-}:&& h = j = \frac{k}{2},~~~~~~~~\,j^{(\I)}= \frac{k+1}{4},~~~j^{(\II)}= \frac{k-1}{4}\nn
d^{+-}_{-\frac{1}{2}}\sigma_{k}^{-}:&& h = j = \frac{k}{2},~~~~~~~~\,j^{(\I)}= \frac{k-1}{4},~~~j^{(\II)}= \frac{k+1}{4}\nn
\sigma_{k}^{+}=d^{++}_{-\frac{1}{2}}d^{+-}_{-\frac{1}{2}}\sigma_{k}^{-}:&& h = j = \frac{k+1}{2},~~~j^{(\I)}= \frac{k+1}{4},~~~j^{(\II)}= \frac{k+1}{4}
\eea
We have also written down the type I and type II charges of each chiral primary. From the construction in \cite{Lunin:2001pw}, the charges of $\sigma_{k}^{\pm}$ comes equally from the two types. The operator $d^{++}_{-\frac{1}{2}}\sigma_{k}^{-}$ carries an extra $1/2$ charge of type I. The operator $d^{+-}_{-\frac{1}{2}}\sigma_{k}^{-}$ carries an extra $1/2$ charge of type II. Under the partial spectral flow by $\alpha_{\I}$ units for type I and $\alpha_{\II}$ units for type II at $z=0$, an operator with charges $j^{(\I)}$ and $j^{(\II)}$ transforms as
\be
O(z)\to z^{-\alpha_{\I} j^{(\I)}-\alpha_{\II} j^{(\II)}}O(z)
\ee
In the cylinder coordinate $w$, the partial spectral flow at $w=-\infty$ becomes 
\be\label{pflow w}
O(w)\to e^{(-\alpha_{\I} j^{(\I)}-\alpha_{\II} j^{(\II)})w}O(w)
\ee
By inserting the type I and type II charges of the chiral primaries in (\ref{chiral}), the transformations of these chiral primaries can be found correspondingly.

\subsection{The deformation}

Everything we have looked at so far applies only at the `orbifold' point of the CFT. If we eventually want to investigate 
the dual gravity theory we must include a deformation operator which moves us toward the supergravity point in the CFT. For some related works about the deformation see \cite{Avery:2010er,Avery:2010hs,Burrington:2014yia,Carson:2014yxa,Carson:2014xwa,Carson:2014ena,Carson:2016cjj,Carson:2016uwf,Carson:2015ohj,Hampton:2019csz,Guo:2021ybz,Guo:2021gqd}.
In this section we show that partial spectral flow is broken by the deformation operator.

The deformed action is 
\be
S=S_0+\lambda \int d^2 w D
\ee
where $S_0$ is the free field action and the deformation operator is given by
\be\label{D 1/4}
D=\frac{1}{4}\epsilon^{\dot A\dot B}\epsilon_{\alpha\beta}\epsilon_{\bar\alpha \bar\beta} G^{\alpha}_{\dot A, -\h} \bar G^{\bar \alpha}_{\dot B, -\h} \sigma^{\beta \bar\beta}_2(w,\bar w)
=\epsilon^{\dot A\dot B} G^{-}_{\dot A, -\h} \bar G^{-}_{\dot B, -\h} \sigma^{++}_2(w,\bar w) 
\ee
where $\epsilon_{+-}=1$ and $\epsilon^{+-}=-1$. The cylinder coordinate $w$ is defined through $z=e^w$.
The twist operator, $\sigma^{++}_2$ factorizes into left and right moving components
\bea
\sigma^{++}_2(w,\bar w)=\s^{+}_2(w)\sigma^{+}_2(\bar w)
\eea
Therefore we only consider the left moving part as the right moving part will be similar. 
The dimension and charge of the $\s^{+}_2$ is $h=j=1/2$ and $j^{(\I)}=j^{(\II)}=1/4$.\footnote{The $\s^{+}_2$ we use in the deformation is actually $\s^{-}_2$ in the first line of (\ref{chiral}). The reason we use that notation is to clearly indicate that the charge is $j=1/2$.}

To find the transformation of the deformation under partial spectral flow, let us first consider the operator $G^{-}_{\dot A,-\h}\sigma^{+}_2(w)$
\be\label{s G}
G^{-}_{\dot A,-\h}\sigma^{+}_2(w)=(G^{-(\I)}_{\dot A,-\h}+G^{-(\II)}_{\dot A,-\h})\sigma^{+}_2(w)
\ee
where we have split the supercharges into type I and type II depending on whether it contains type I fermions or type II fermions as in (\ref{G++}) and (\ref{G+-}). The $G^{-(\I)}_{\dot A,-\h}$ has $-1/2$ unit of type I charge and $G^{-(\II)}_{\dot A,-\h}$ has $-1/2$ unit of type II charge. Thus we have the charges
\bea\label{G 12 charge}
G^{-(\I)}_{\dot A,-\h}\sigma^{+}_2(w): &&j=0,~~~j^{(\I)}=-\frac{1}{4},~~~j^{(\II)}=\frac{1}{4}\nn
G^{-(\II)}_{\dot A,-\h}\sigma^{+}_2(w): &&j=0,~~~j^{(\I)}=\frac{1}{4},~~~\,~~j^{(\II)}=-\frac{1}{4}
\eea

The transformation of partial spectral flow (\ref{pflow w}) gives
\be
G^{-}_{\dot A,-\h}\sigma^{+}_2(w)
\rightarrow \big(e^{\frac{1}{4}(\alpha_{\I}-\alpha_{\II}) w}G^{-(\I)}_{\dot A,-\h}+e^{-\frac{1}{4}(\alpha_{\I}-\alpha_{\II}) w}G^{-(\II)}_{\dot A,-\h}\big)\sigma^{+}_2(w)
\ee
As we can see, the deformation operator is not generally invariant under partial spectral flow. Only when $\alpha_{\I}=\alpha_{\II}$, which corresponds to the original spectral flow or at $w=0$, is it invariant. Therefore, partial spectral flow is broken as one perturbs away from the orbifold point by adding the deformation (\ref{D 1/4}). In the next section, we will study its effect on the spectrum of D1D5P states.

\section{Second-order energy lift}
\label{lifting}

\subsection{The Gava-Narain method}

In this subsection we briefly review the method of computing the second-order energy lift of D1D5P states. It was proposed by Gava and Narain \cite{Gava:2002xb} and further investigated in \cite{Guo:2019pzk}. For other works about lifting of 
states, see \cite{Gaberdiel:2015uca,Hampton:2018ygz,Guo:2019ady,Guo:2020gxm,Lima:2020boh,Lima:2020kek,Lima:2020nnx,Lima:2020urq,Lima:2021wrz,Benjamin:2021zkn}. We will also show that states related by spectral flow have the same amount of lift. In the next subsection we will show that it is not true for partial spectral flow.

We will work in the Ramond sector. At the orbifold point, there are Ramond ground states with dimensions $h=\bar h=\frac{c}{24}$, which are called D1D5 states. If we only excite the left movers, the states obtained are called D1D5P states. They have dimensions
\be\label{dimen}
(h,\bar h)=(n+\frac{c}{24},\frac{c}{24})
\ee
where $n$ is an integer. This $n$ is also called the `level' of states.
As we move away from the orbifold point, the dimensions of the D1D5P states can change
\be
(h,\bar h)\to (h+\delta h, \bar h + \delta \bar h)
\ee
The spin $h-\bar h$ must be an integer which implies that the small change in the left and right moving dimensions must be the same
\be
\delta h = \delta \bar h
\ee
Furthermore, because Ramond ground states have the lowest possible dimension, the change in dimension for the right mover must be non-negative, $\delta \bar h\geq 0$. Therefore the energy lift must be non-negative
\be\label{E p}
E^{(2)}=2 \delta h \geq 0
\ee
This non-negative result implies that the first nontrivial contribution starts at second order in the perturbation $\lambda$, where $\delta h \sim O(\lambda^2)$.

Gava and Narain proposed a method to compute this energy lift for D1D5P states. Consider the subspace formed by unperturbed states $O^{(0)}_{a}$ at level-$n$, which have dimension (\ref{dimen}). In this subspace, define the lifting matrix
\bea\label{liftmatrix 1}
E^{(2)}_{ba}=2 \lambda^2   
\Big\langle O^{(0)}_{b}\Big|    \Big\{  \bar G^{+(P)\dagger}_{+,0},  \bar G^{+(P)}_{+,0} \Big\} \Big|O^{(0)}_{a}\Big\rangle
=2 \lambda^2   
\Big\langle O^{(0)}_{b}\Big|    \Big\{  \bar G^{+(P)\dagger}_{-,0},  \bar G^{+(P)}_{-,0} \Big\} \Big|O^{(0)}_{a}\Big\rangle
\eea
where
\be\label{GN p s}
\bar G^{\bar \alpha (P)}_{\dot A,0}
= \pi \mathcal P G^{+}_{\dot A,-\frac{1}{2}}\sigma^{-\bar \alpha} \mathcal P
= - \pi \mathcal P G^{-}_{\dot A,-\frac{1}{2}}\sigma^{+\bar \alpha} \mathcal P
\ee
where the operator of the form $G\sigma$ is sitting at $w=0$ and the projection operator $\mathcal P$ projects any state to the level-$n$ subspace. It has been shown that $\bar G^{\bar \alpha (P)}_{\dot A,0}$ is related to the first order correction to the supercharge $\bar G^{\bar \alpha}_{\dot A,0}$. We note that there is some redundancy in the projection operator when we insert (\ref{GN p s}) into (\ref{liftmatrix 1}).
The Hermitian conjugation relations are
\be
\bar G^{+ (P)\dagger}_{+,0} = - \bar G^{- (P)}_{-,0},~~~~~\bar G^{+ (P)\dagger}_{-,0} =  \bar G^{- (P)}_{+,0}
\ee
The eigenvalue of the lifting matrix $E^{(2)}$ is the value of the lift. The corresponding eigenstate is the zeroth order state having this lift.

In the following, we will be interested in the average lift for a state $\Phi$ at level-$n$
\be\label{lift average}
E^{(2)}(\Phi)
=2\lambda^2\pi^2\langle \Phi|\!\left[( G^{+}_{-,-\h}\sigma^{--})P_{h(\Phi)}( G^{-}_{+,-\h}\sigma^{++})+( G^{+}_{+,-\h}\sigma^{-+})P_{h(\Phi)}( G^{-}_{-,-\h}\sigma^{+-})\right]\!|\Phi\rangle
\ee
where the projection operator $P_{h(\Phi)}$ projects states onto the subspace of states having the same dimension as the state $\Phi$.
Because eigenvalues of $E^{(2)}$ are non-negative, if the above average is positive, the state $\Phi$ must contain a lifted state.

\subsection{No lift from spectral flow}

Spectral flow on the left mover will map a D1D5P state to another D1D5P state. In this section we will show that the second-order energy lift of D1D5P states is invariant under spectral flow on the left mover. 
Under spectral flow, we have
\be
G^{-}_{\dot A,-\h}\sigma^{+} \to G^{-}_{\dot A,-\h}\sigma^{+}
\ee
since the operator has zero charge. Let us take the spectral flow of (\ref{lift average}).
We will use $'$ to label the state after spectral flow.
We obtain 
\be
E^{(2)}(\Phi)=2\lambda^2\pi^2\langle \Phi'|\!\left[( G^{+}_{-,-\h}\sigma^{--})P'_{h(\Phi)}( G^{-}_{+,-\h}\sigma^{++})+( G^{+}_{+,-\h}\sigma^{-+})P'_{h(\Phi)}( G^{-}_{-,-\h}\sigma^{+-})\right]\!|\Phi'\rangle
\ee
Before spectral flow the intermediate state obtained from the projection operator $P_{h(\Phi)}$ has the same charge and dimension as the initial state $\Phi$. After spectral flow the intermediate state and the initial state $\Phi'$ will also have the same charge and dimension. Thus
\be
P'_{h(\Phi)} = P_{h(\Phi')}
\ee
Then we have
\bea
E^{(2)}(\Phi)
\!\!&=&\!\!2\lambda^2\pi^2\langle \Phi'|\!\left[( G^{+}_{-,-\h}\sigma^{--})P_{h(\Phi')}( G^{-}_{+,-\h}\sigma^{++})+( G^{+}_{+,-\h}\sigma^{-+})P_{h(\Phi')}( G^{-}_{-,-\h}\sigma^{+-})\right]\!|\Phi'\rangle\nn
\!\!&=&\!\!E^{(2)}(\Phi')
\eea
The states $\Phi$ and $\Phi'$ have the same lift. Spectral flow does not change the value of the lift. 

\subsection{Lift from partial spectral flow}

In this subsection, we will show that the lift will change in general under partial spectral flow.
For partial spectral flow, let us separate the operator based on its type I and type II charges as in (\ref{s G})
\be
G^{-}_{\dot A,-\h}\sigma^{+} = G^{-(\I)}_{\dot A,-\h}\sigma^{+} + G^{-(\II)}_{\dot A,-\h}\sigma^{+}
\ee
From (\ref{G 12 charge}), the charges are 
\bea\label{G 12 charge 1}
G^{-(\I)}_{\dot A,-\h}\sigma^{+} ~:~~~j^{(\I)}=-\frac{1}{4},~~~j^{(\II)}=\frac{1}{4}\nn
G^{-(\II)}_{\dot A,-\h}\sigma^{+} ~:~~~j^{(\I)}=\frac{1}{4},~~~j^{(\II)}=-\frac{1}{4}
\eea
The lift (\ref{lift average}) becomes
\bea\label{lift 12}
E^{(2)}(\Phi)&=&2\lambda^2\pi^2\langle \Phi|\Big[( G^{+(\I)}_{-,-\h}\sigma^{--})P_{h(\Phi),j^{(\I)}(\Phi)-\frac{1}{4},j^{(\II)}(\Phi)+\frac{1}{4}}( G^{-(\I)}_{+,-\h}\sigma^{++})\nn
&&~~~~~~\,~~~~+( G^{+(\II)}_{-,-\h}\sigma^{--})P_{h(\Phi),j^{(\I)}(\Phi)+\frac{1}{4},j^{(\II)}(\Phi)-\frac{1}{4}}( G^{-(\II)}_{+,-\h}\sigma^{++})\nn
&&~~~~~~\,~~~~+( G^{+(\I)}_{+,-\h}\sigma^{-+})P_{h(\Phi),j^{(\I)}(\Phi)-\frac{1}{4},j^{(\II)}(\Phi)+\frac{1}{4}}( G^{-(\I)}_{-,-\h}\sigma^{+-})\nn
&&~~~~~~\,~~~~+( G^{+(\II)}_{+,-\h}\sigma^{-+})P_{h(\Phi),j^{(\I)}(\Phi)+\frac{1}{4},j^{(\II)}(\Phi)-\frac{1}{4}}( G^{-(\II)}_{-,-\h}\sigma^{+-})
\Big]|\Phi\rangle
\eea
where we have written type I and type II charges of the intermediate states explicitly. 
Because type I and type II charges are conserved there is no cross term between these two types.

Let us now consider the transformation of the projection operators in (\ref{lift 12})
\be\label{two p}
P_{h(\Phi),j^{(\I)}(\Phi)-\frac{1}{4},j^{(\II)}(\Phi)+\frac{1}{4}},
~~~~P_{h(\Phi),j^{(\I)}(\Phi)+\frac{1}{4},j^{(\II)}(\Phi)-\frac{1}{4}}
\ee
For an intermediate state with dimension and charges required by the first one, it transforms to a state with dimension
\bea
h' &=& h(\Phi)+\alpha_{\I}(j^{(\I)}(\Phi)-\frac{1}{4})+\alpha_{\II}(j^{(\II)}(\Phi)+\frac{1}{4})
+\frac{c}{48}(\alpha_{\I}^2+\alpha_{\II}^2)\nn
&=& h(\Phi') - \frac{1}{4}(\alpha_{\I}-\alpha_{\II})
\eea
Thus we have
\be
P_{h(\Phi),j^{(\I)}(\Phi)-\frac{1}{4},j^{(\II)}(\Phi)+\frac{1}{4}} \to P_{h(\Phi')- \frac{1}{4}(\alpha_{\I}-\alpha_{\II})}
\ee
where we do not write charges of the resulting projection explicitly because it will be satisfied automatically.
Similarly, we have
\be
P_{h(\Phi),j^{(\I)}(\Phi)+\frac{1}{4},j^{(\II)}(\Phi)-\frac{1}{4}} \to P_{h(\Phi')+ \frac{1}{4}(\alpha_{\I}-\alpha_{\II})}
\ee
Notice that since operators of the form $G\sigma$ in (\ref{lift 12}) is sitting at $w=0$ they are unchanged under partial spectral flow.
Taking partial spectral flow of (\ref{lift 12}), we obtain
\bea\label{lift 12 flow}
E^{(2)}(\Phi)&=&2\lambda^2\pi^2\langle \Phi'|\Big[( G^{+(\I)}_{-,-\h}\sigma^{--})
P_{h(\Phi')- \frac{1}{4}(\alpha_{\I}-\alpha_{\II})}
( G^{-(\I)}_{+,-\h}\sigma^{++})\nn
&&~~~~~~~~~~~+( G^{+(\II)}_{-,-\h}\sigma^{--})
P_{h(\Phi')+ \frac{1}{4}(\alpha_{\I}-\alpha_{\II})}
( G^{-(\II)}_{+,-\h}\sigma^{++})\nn
&&~~~~~~~~~~~+( G^{+(\I)}_{+,-\h}\sigma^{-+})
P_{h(\Phi')- \frac{1}{4}(\alpha_{\I}-\alpha_{\II})}
( G^{-(\I)}_{-,-\h}\sigma^{+-})\nn
&&~~~~~~~~~~~+( G^{+(\II)}_{+,-\h}\sigma^{-+})
P_{h(\Phi')+ \frac{1}{4}(\alpha_{\I}-\alpha_{\II})}
( G^{-(\II)}_{-,-\h}\sigma^{+-})
\Big]|\Phi'\rangle\nn
&\neq& E^{(2)}(\Phi')
\eea
where $E^{(2)}(\Phi')$ is given by (\ref{lift 12}) with $\Phi$ replaced by $\Phi'$. There is only the projection operator $P_{h(\Phi')}$ but not $P_{h(\Phi')\pm \frac{1}{4}(\alpha_{\I}-\alpha_{\II})}$. 
This difference in projection operators makes the lift of $\Phi$ and $\Phi'$ unequal. This is due to the fact that the operators of the form $G\sigma$ in (\ref{G 12 charge 1}) have nonzero type I and type II charges even though the total charge is zero.
In the next section, we will give an explicit example.

\subsection{An example}\label{sec example}

In this subsection, we will give an explicit example to show that partial spectral flow changes the lift of states. 
We will consider a state $\Phi$ of two singly wound strings in the R sector
\be\label{state}
|\Phi\rangle = |\Phi_L\rangle |\Phi_R\rangle
\ee
where the right mover is
\be\label{r}
|\Phi_R\rangle = |\bar 0_R^-\rangle^{(1)} |\bar 0_R^-\rangle^{(2)}
\ee
where we use the notation $0^-_R$ ($\bar 0^-_R$) to denote the Ramond ground state of a singly wound string with charge $-{1\over2}$ for left (right) movers respectively. Here the two singly wound strings are labelled by copy 1 and copy 2.
For the sector of two singly wound strings, it has been shown in \cite{Guo:2019ady,Guo:2020gxm} that only states with a symmetric right mover can have nonzero lift. Thus we take the right mover (\ref{r}) which is symmetric under the replacement of the copy labels $(1)\leftrightarrow (2)$. As required by the orbifold theory, the state (\ref{state}) with both left and right movers must be symmetric between copy 1 and copy 2. Since the right mover (\ref{r}) is symmetric the left mover must be symmetric.
Thus we take the following symmetric left mover
\be\label{l}
|\Phi_L\rangle=d^{++(1)}_{-(2n-1)}d^{++(2)}_{-(2n-1)}\ldots d^{++(1)}_{-1}d^{++(2)}_{-1}d^{++(1)}_0d^{++(2)}_0| 0_R^-\rangle^{(1)} | 0_R^-\rangle^{(2)}
\ee
where $n$ is a positive integer. This left mover is symmetric because there are an even number of fermions on each copy.
This state (\ref{l}) fills the fermi sea of the type I fermion $d^{++}$ but it is empty for the type II fermion $d^{+-}$.
Let us take a partial spectral flow with $\alpha_{\I}= - 4n$. The state $\Phi$ becomes 
\be
|\Phi\rangle \to |\Phi'\rangle=| 0_R^-\rangle^{(1)} | 0_R^-\rangle^{(2)} |\bar 0_R^-\rangle^{(1)} |\bar 0_R^-\rangle^{(2)}
\ee
Since $\Phi'$ is a Ramond ground state, it is unlifted. In the following, we will compute the lift of $\Phi$. We will show that the lift is nonzero. Thus states related by partial spectral flow do not have the same lift in general.

First we use the following relations, which have been shown in \cite{Guo:2019ady,Guo:2020gxm}, to simplify the right movers of (\ref{lift 12 flow}) 
\be
{}^{(2)}\langle \bar 0_R^- |{}^{(1)}\langle \bar 0_R^- |  \bar \sigma^{-} P \bar \sigma^{+} |\bar 0_R^-\rangle^{(1)} |\bar 0_R^-\rangle^{(2)} = 1
\ee
and
\be
{}^{(2)}\langle \bar 0_R^- |{}^{(1)}\langle \bar 0_R^- |  \bar \sigma^{+} P \bar \sigma^{-} |\bar 0_R^-\rangle^{(1)} |\bar 0_R^-\rangle^{(2)} = 0
\ee
where $P$ projects onto the right moving Ramond ground states.
Thus (\ref{lift 12 flow}) becomes
\bea
E^{(2)}(\Phi)&=&-2\lambda^2\pi^2\langle \Phi'_L|\big[( G^{+(\I)}_{-,-\h}\sigma^{-})
P_{h(\Phi')+ n}
( G^{-(\I)}_{+,-\h}\sigma^{+})\nn
&&~~~~~~~~~~~~+( G^{+(\II)}_{-,-\h}\sigma^{-})
P_{h(\Phi')- n}
( G^{-(\II)}_{+,-\h}\sigma^{+})
\big]|\Phi'_L\rangle
\eea
where we have included an extra minus sign due to the fact that when we combine the left and right movers there is a minus sign from the Hermitian conjugation relation $(\sigma^{++})^{\dagger}= -\sigma^{--}$.
The left mover, $\Phi'_L$, is given by
\be
|\Phi'_L\rangle=| 0_R^-\rangle^{(1)} | 0_R^-\rangle^{(2)}
\ee
Because $\Phi'$ is a Ramond ground state, it has the lowest possible dimension. In the second term, there is no state with dimension $h(\Phi')- n$ for $n>0$ as required by the projection operator. Thus only the first term is nonzero. We have
\be\label{example 1}
E^{(2)}(\Phi)=-2\lambda^2\pi^2
\left[{}^{(2)}\langle 0_R^- | {}^{(1)}\langle 0_R^-|
( G^{+(\I)}_{-,-\h}\sigma^{-})
P_{h(\Phi')+ n}
( G^{-(\I)}_{+,-\h}\sigma^{+})| 0_R^-\rangle^{(1)} | 0_R^-\rangle^{(2)}\right]
\ee
To compute this value of the lift, let us first look at the following amplitude derived in \cite{Hampton:2018ygz}
\be\label{G base}
{}^{(2)}\langle 0_R^- | {}^{(1)}\langle 0_R^-|
( G^{+}_{-,-\h}\sigma^{-}(w_2))
( G^{-}_{+,-\h}\sigma^{+}(w_1))
| 0_R^-\rangle^{(1)} | 0_R^-\rangle^{(2)}
=-\frac{1}{4\sinh^2(\frac{\Delta w}{2})}
\ee
where $\Delta w= w_2 - w_1$. Similarly we have 
\bea\label{amp}
{}^{(2)}\langle 0_R^- | {}^{(1)}\langle 0_R^-|
( G^{+(\I)}_{-,-\h}\sigma^{-}(w_2))
( G^{-(\I)}_{+,-\h}\sigma^{+}(w_1))
| 0_R^-\rangle^{(1)} | 0_R^-\rangle^{(2)}
\!&=&\!-\frac{1}{8\sinh^2(\frac{\Delta w}{2})}\nn
\!&=&\! - \frac{1}{2}\sum_{k> 0}k\, e^{-k\Delta w}
\eea
Here since there is only a type I supercharge, we obtain half of the contribution compared to (\ref{G base}). 

The projection $P_{h(\Phi')+ n}$ in (\ref{example 1}) requires us to pick the term $e^{-n\Delta w}$, where $n$ is a positive integer.
This gives
\be
k=n
\ee
In (\ref{amp}), the corresponding term has the coefficient 
\be\label{lift}
 {}^{(2)}\langle 0_R^- |{}^{(1)} \langle 0_R^-|
( G^{+(\I)}_{-,-\h}\sigma^{-})
P_{h(\Phi')+ n}
( G^{-(\I)}_{+,-\h}\sigma^{+})| 0_R^-\rangle^{(1)} | 0_R^-\rangle^{(2)}=-\frac{n}{2}
\ee
Thus from (\ref{example 1}) the lift of the state $\Phi$ is
\be\label{lift p}
E^{(2)}(\Phi)= \lambda^2\pi^2 n
\ee
where $n$ is a positive integer. Thus we have shown that the state $\Phi$ and $\Phi'$ which are related by partial spectral flow do not have the same lift.

\subsection{Comparison with previous lifting results}

In this subsection we will compare the result of lifting obtained in the previous subsection with  some results in \cite{Hampton:2018ygz,Guo:2019pzk,Guo:2019ady}. 

Let us first compare to \cite{Guo:2019pzk,Guo:2019ady}. In these works, the lifting of D1D5P states for two singly wound strings has been studied up to level-$4$. 
There are three primaries at level-1. They can be organized into a triplet of charge $A$, which is carried by the $A$ indices in modes $d^{\alpha A}_{-n}$ and $\alpha_{A\dot A,-n}$. They have the same lift
\be\label{1 lift}
E^{(2)}_{\text{level}-1} = \lambda^2 \pi^2
\ee
The top member of the triplet is 
\be\label{1 pr}
\frac{1}{2}(d^{-+(1)}_{-1}-d^{-+(2)}_{-1})(d^{++(1)}_{0}-d^{++(2)}_{0})| 0_R^-\rangle^{(1)} | 0_R^-\rangle^{(2)}|\bar 0_R^-\rangle^{(1)} |\bar 0_R^-\rangle^{(2)}
\ee
By applying $(d^{++(1)}_{-1}+d^{++(2)}_{-1})(d^{++(1)}_{0}+d^{++(2)}_{0})J^+_0$ to the above state and normalizing the resulting state, we obtain the state (\ref{state}) with $n=1$
\be
d^{++(1)}_{-1}d^{++(2)}_{-1}d^{++(1)}_0d^{++(2)}_0| 0_R^-\rangle^{(1)} | 0_R^-\rangle^{(2)}
|\bar 0_R^-\rangle^{(1)} |\bar 0_R^-\rangle^{(2)}
\ee
Thus the above state, which is at level-2, is a descendent of the level-1 primary (\ref{1 pr}). Therefore, the above state is an eigenstate of the lifting matrix and has the lift (\ref{1 lift}).
As we can see it is consistent with the lift (\ref{lift p}) of the $n=1$ state derived by partial spectral flow.

\begin{figure}
\centering
        \includegraphics[width=16cm]{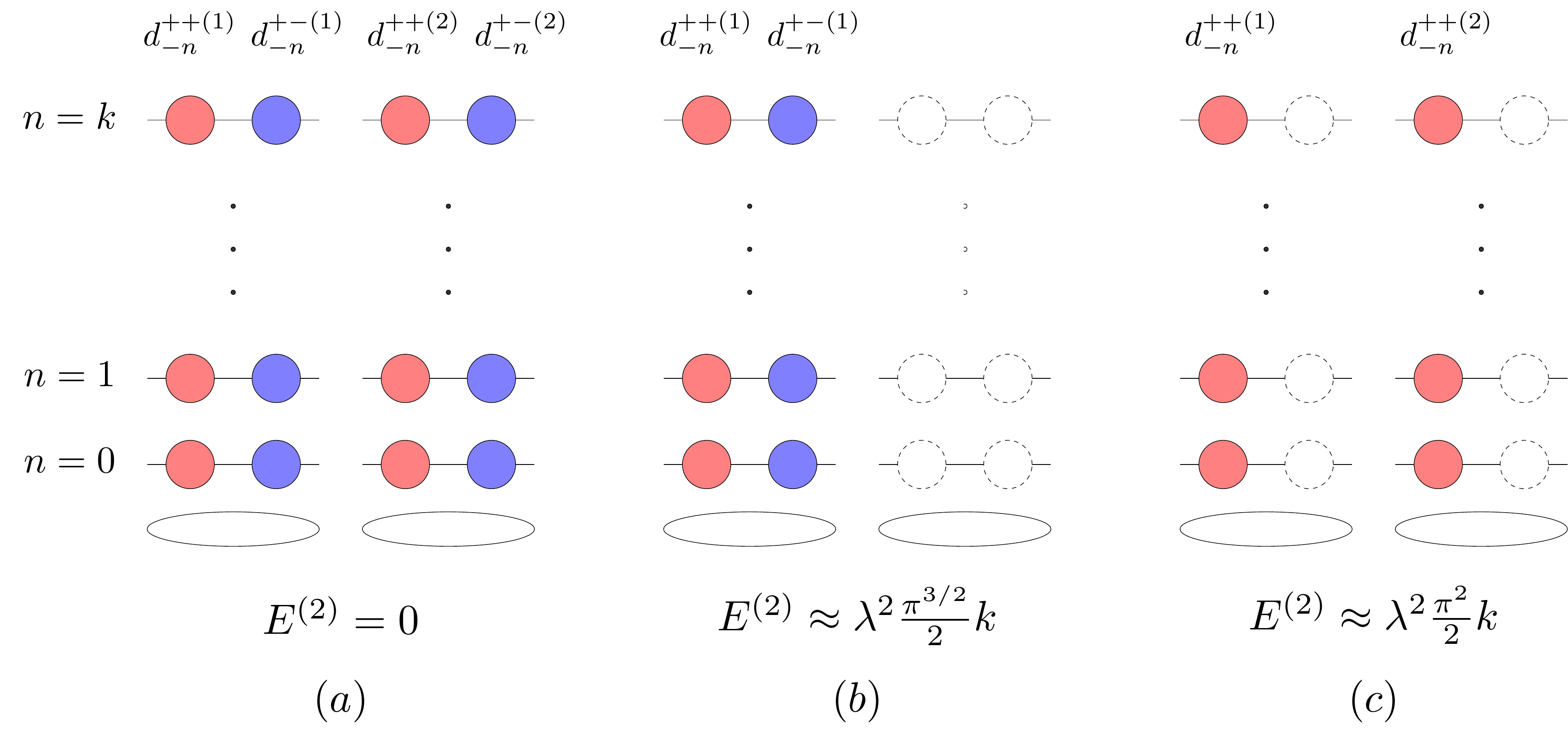}
\caption{Various fillings of the fermi sea and their lifts.}
\label{fig}
\end{figure}

Then let us compare to \cite{Hampton:2018ygz}, where the relation between lifting and spectral flow has been studied. If we spectral flow both singly wound strings, we obtain the following state
\be\label{f 1}
d^{++(1)}_{-k}d^{+-(1)}_{-k}d^{++(2)}_{-k}d^{+-(2)}_{-k}\dots
 d^{++(1)}_0d^{+-(1)}_0d^{++(2)}_0d^{+-(2)}_0
| 0_R^-\rangle^{(1)} | 0_R^-\rangle^{(2)}|\bar 0_R^-\rangle^{(1)} |\bar 0_R^-\rangle^{(2)}
\ee
The fermi sea is filled for both type I and type II fermions and for both strings, which is depicted  in fig.\ref{fig}(a).
It has zero lift. Now let us spectral flow only one of the two singly wound strings. In this way, we obtain the following state
\be\label{sea 1}
d^{++(1)}_{-k}d^{+-(1)}_{-k}
\ldots d^{++(1)}_0d^{+-(1)}_0
| 0_R^-\rangle^{(1)} | 0_R^-\rangle^{(2)}|\bar 0_R^-\rangle^{(1)} |\bar 0_R^-\rangle^{(2)}
\ee
The fermi sea is filled for both type I and type II fermions but only for the first string,
which is depicted in fig.\ref{fig}(b). The lift has been found in \cite{Hampton:2018ygz}, which is
\be
E^{(2)} = \lambda^2 \frac{\pi^{3/2}}{2}\frac{\Gamma[k^2+k-\frac{1}{2}]}{\Gamma[k^2+k-1]} \approx \lambda^2 \frac{\pi^{3/2}}{2} k
\ee
where in the last step we have taken the large $k$ limit.
To make the comparison with the result in the previous subsection, let us define $k=2n-1$. The state (\ref{state}) becomes 
\be\label{sea 2}
|\Phi\rangle=d^{++(1)}_{-k}d^{++(2)}_{-k}\ldots d^{++(1)}_0d^{++(2)}_0| 0_R^-\rangle^{(1)} | 0_R^-\rangle^{(2)}|\bar 0_R^-\rangle^{(1)} |\bar 0_R^-\rangle^{(2)}
\ee
where $k$ is an odd number. The fermi sea is filled for only type I fermions and for both strings, which is depicted  in fig.\ref{fig}(c). The lift (\ref{lift p}) becomes
\be
E^{(2)}(\Phi)= \lambda^2\frac{\pi^2}{2} (k+1) \approx \lambda^2\frac{\pi^2}{2} k
\ee
where in the last step we have taken the large $k$ limit. Thus we can see that when the fermi sea is filled completely, the lift is zero. When the fermi sea is filled halfway either in (\ref{sea 1}) or (\ref{sea 2}), the lift is proportional to $k$ in the large $k$ limit, where $k$ is the energy of the fermi surface.

In the following, we give an interpretation of why states related by spectral flow have the same lift while states related by partial spectral flow do not. Consider the state depicted in fig.\ref{fig}(a). Applying the Gava-Narain method directly, the number of possible intermediate states in (\ref{lift average}) grows with the energy of the fermi surface. Thus the zero lift of the state is not caused simply by the naive idea that because the fermi sea is filled, there are no possible intermediate states which contribute.
This suggests that the zero lift of this state is actually caused by a cancellation between intermediate states coming from initial modes living on both copies and consisting of both types. However for the state depicted in fig.\ref{fig}(b), the cancellation between modes living on copy 1 and copy 2 is missing.
While for the state given in fig.\ref{fig}(c), the cancellation between modes consisting of type I and type II is missing. Thus the states depicted in fig.\ref{fig}(b) and fig.\ref{fig}(c) are lifted.

\section{Conclusion}\label{conclusion}

In this work we have studied an enhanced automorphism for the orbifold (free) point of the D1D5 CFT. We call it partial spectral flow. At the orbifold point, the $\mathcal{N}=4$ superconformal algebra (SCA) is realized by using four free fermions and four free bosons. The four fermions are grouped into two types
\bea
\text{Type I}: ~~d^{++}_{r},~~~ d^{--}_{r}\cr
\text{Type II}: ~~ d^{+-}_{r},~~~ d^{-+}_{r}
\eea
which decouple from one another. Partial spectral flow flows the two types of fermions separately. 
We showed that partial spectral flow keeps the $\mathcal{N}=4$ SCA invariant and is thus a valid automorphism at the orbifold point.  
The grouping allowed some generators of the $\mathcal{N}=4$ algebra namely 
$L_n,J^3_n, G^{+}_{+,r},G^{-}_{-,r}$ \footnote{Or generators $L_n,J^3_n, G^{+}_{-,r},G^{-}_{+,r}$. See subsection \ref{reduction to subalgebra} for more details.} to also be decomposed into two types. These two types of generators close within each type and form two decoupled $\mathcal{N}=2$ algebras. We have shown that partial spectral flow is indeed the two spectral flows of these two $\mathcal{N}=2$ algebras. 

In the orbifold theory, there exist twist operators that can link several copies into a single multi-wound copy. We showed that these twist operators have well-defined transformation properties under partial spectral flow. To move away from the orbifold point, we need to include the deformation operator.
Using the transformation property of the twist operator we also found the transformation of the deformation operator. Partial spectral flow does not leave the deformation operator invariant generically. It is broken by the deformation operator.

We studied the effect of this breaking on the spectrum of D1D5P states. When the theory is deformed away from the orbifold point, the energy lift of the D1D5P states begin at second order in perturbation theory. We have shown that D1D5P states related by partial spectral flow do not have the same energy lift in general. The lift of a state can be computed by its partial spectral flowed state through (\ref{lift 12 flow}). Based on this result, we computed the lift for a class of states in the two singly wound sector. This state is obtained by applying partial spectral flow to a Ramond ground state. Only half of the fermi sea is filled as depicted in fig.\ref{fig}(c). The lift is proportional to $k$ in the large $k$ limit, where $k$ is the energy of the fermi surface.

In the following let us give a list of comments and future directions:

\b

(i) At the orbifold point, the symmetry is actually much larger than the $\mathcal{N}=4$ superconformal symmetry. It contains higher spin symmetry which comes from the tensionless string in AdS \cite{Prokushkin:1998bq,Vasiliev:2003ev,Gaberdiel:2014cha,Gaberdiel:2015mra}.
In section \ref{reduction to subalgebra}, we obtained two types of generators from decomposing the $\mathcal{N}=4$ generators. These new generators, which are bilinear in the free fields, are indeed the generators of the higher spin symmetry \cite{Gaberdiel:2015uca}. The higher spin symmetry comes from the basic commutators (\ref{basic com}). Thus partial spectral flow, which preserves the basic commutators, is also an automorphism of the higher spin symmetry.

\b

(ii) At the orbifold point, there exist even more general spectral flows than partial spectral flow. For example, one can spectral flow even modes of the type I fermions $d^{++}_{n}$ and $d^{--}_{n}$ by $\alpha_{\I,\text{even}} = 4m$, where $m$ is an integer. In this way, only even modes of the type I fermions are flowed into each other to keep the basic commutator (\ref{basic com}). This is an automorphism at the orbifold point. However, the twist operators do not have well-defined transformation properties under these flows since they mix the even and odd modes. As a result, the deformation operator, which contains a twist operator, does not have a well-defined transformation. Thus away from the orbifold point, the effect of breaking of these more general spectral flows can not be studied easily compared to partial spectral flow. It would be interesting to find a way to further generalize partial spectral flow while still having a relatively simply transformation of the deformation operator.

\b

(iii) Even though partial spectral flow is broken away from the orbifold point, the effect of the breaking can still be studied due to the well-defined transformation property of the deformation operator. The orbifold point corresponds to tensionless strings in AdS, while adding a deformation corresponds to adding tension to the strings. It would be interesting to see if partial spectral flow could be a useful probe for exploring this physics.

\b

(iv) The example given in subsection \ref{sec example} shows that states related by partial spectral flow do not have the same lift. 
It is also an example of how a complicated twist operator computation can be simplified by using partial spectral flow. 
It would be interesting to see if partial spectral flow could be used to simplify other computations involving the twist operator.

\section{Acknowledgements}
We would like to thank Zaq Carson and Samir D. Mathur for their early collaboration on this project. We would also like to thank Samir D. Mathur for helpful discussions. The work of B.G. is supported by the DOE grant DE-SC0011726 and the ERC Grant 787320 - QBH Structure. The work of S.H. is supported by the ERC Grant 787320 - QBH Structure.

\appendix

\section{Partial spectral flow of the $\mathcal{N}=4$ superconformal algebra}

The $\mathcal{N}=4$ SCA is invariant under partial spectral flow (\ref{flow 1}) and (\ref{flow 2}). In this appendix, we will give some examples. We will mainly focus on type I partial spectral flow (\ref{flow 1}) for commutators with terms containing the central charge.

\b

(i) Consider the commutator
\bea
[J^+_{m},J^-_{n}]={c\over6}m\delta_{m+n,0} + 2J^3_{m+n}
\label{jpjm}
\eea
Partial spectral flowing the left-hand side (LHS) of (\ref{jpjm}) by $\a_\I$ yields
\bea\label{1l}
\text{LHS}&\to&[J^+_{m+{\alpha_{\I}\over 2}},J^-_{n-{\alpha_{\I}\over 2}}]
=\frac{c}{6}(m+\frac{\alpha_{\I}}{2})\delta_{m+n}+2 J^3_{m+n}
\eea
The right-hand side (RHS) partial spectral flows to
\bea\label{1r}
\text{RHS}&\to&{c\over6}m\delta_{m+n,0} + 2(J^3_{m+n}+{\a_\I c\over24}\delta_{m+n,0})=\frac{c}{6}(m+\frac{\alpha_{\I}}{2})\delta_{m+n}+2 J^3_{m+n}
\eea
We can see that (\ref{1l}) and (\ref{1r}) are equal.

\b

(ii) Consider the anticommutator
\bea
\lbrace G^{+}_{+,r} , G^{-}_{-,s} \rbrace=  -\bigg[{c\over6}(r^2-{1\over4})\delta_{r+s,0}  +(r-s)J^3_{r+s}  + L_{r+s}  \bigg]
\label{gpgm}
\eea
Partial spectral flowing the LHS of (\ref{gpgm}) by $\a_\I$ yields
\bea
\text{LHS}&\to&\lbrace G^{+(\I)}_{+,r+\alpha_{\I}/2} +G^{+(\II)}_{+,r}, G^{-(\I)}_{-,s-\alpha_{\I}/2}+G^{-(\II)}_{-,s} \rbrace\cr
&=&\lbrace G^{+(\I)}_{+,r+\alpha_{\I}/2} , G^{-(\I)}_{-,s-\alpha_{\I}/2} \rbrace
+\lbrace G^{+(\II)}_{+,r},G^{-(\II)}_{-,s}\rbrace
\label{GG anticommutator}
\eea
where
\bea
\lbrace G^{+(\I)}_{+,r+\alpha_{\I}/2} , G^{-(\I)}_{-,s-\alpha_{\I}/2} \rbrace
&=&-\bigg[{c/2\over6}((r+\alpha_{\I}/2)^2-{1\over4})\delta_{r+s,0}  +(r-s+\a_\I)J^{3(\I)}_{r+s}  + L^{(\I)}_{r+s}\bigg]\cr
&=&-\bigg[{c/2\over6}(r^2-{1\over4})\delta_{r+s,0}  +(r-s)(J^{3(\I)}_{r+s}+\frac{\alpha_{\I} c}{24}\delta_{r+s,0}) \cr
&&\qquad + L^{(\I)}_{r+s}+\a_\I  J^{3(\I)}_{r+s} +  {\a_\I^2c\over48}\delta_{r+s,0} \bigg]\cr
\cr
\lbrace G^{+(\II)}_{+,r} , G^{-(\II)}_{-,s} \rbrace&=&  -\bigg[{c/2\over6}(r^2-{1\over4})\delta_{r+s,0}  +(r-s)J^{3(\II)}_{r+s}  + L^{(\II)}_{r+s}  \bigg]
\eea
Thus we have
\bea\label{l 2}
\text{LHS}&\to& -\bigg[{c\over6}(r^2-{1\over4})\delta_{r+s,0}  +(r-s)(J^{3(\I)}_{r+s}+\frac{\alpha_{\I} c}{24}\delta_{r+s,0}+J^{3(\II)}_{r+s}) \cr
&& \qquad+ L^{(\I)}_{r+s}+\a_\I  J^{3(\I)}_{r+s} +  {\a_\I^2c\over48}\delta_{r+s,0}+ L^{(\II)}_{r+s} \bigg]
\eea
The RHS of (\ref{gpgm}) partial spectral flows to
\bea
\text{RHS}&\to&-\bigg[{c\over6}(r^2-{1\over4})\delta_{r+s,0}  +(r-s)(J^{3(\I)}_{r+s}+\frac{\alpha_{\I} c}{24}\delta_{r+s,0} + J^{3(\II)}_{r+s}) \cr
&&\qquad + L^{(\I)}_{r+s}+\a_\I  J^{3(\I)}_{r+s} +  {\a_\I^2c\over48}\delta_{r+s,0} + L^{(\II)}_{r+s}\bigg]
\eea
which equals partial spectral flow of the LHS (\ref{l 2}).

Similarly the anticommutator 
 \bea
 \lbrace G^{+}_{-,r} , G^{-}_{+,s} \rbrace&=& \bigg[{c\over6}(r^2-{1\over4})\delta_{r+s,0}  +(r-s)J^3_{r+s}  + L_{r+s}  \bigg]
\eea
is invariant under partial spectral flow.

\b

(iii) Consider the anticommutator
\bea
\lbrace G^{+}_{+,r} , G^{+}_{-,s} \rbrace=  (r-s)J^+_{r+s}
\label{gpgp}
\eea
The LHS can be written as
\bea
\text{LHS} &=& \lbrace G^{+(\I)}_{+,r} + G^{+(\II)}_{+,r}, G^{+(\I)}_{-,s} + G^{+(\II)}_{-,s} \rbrace
\cr
&=&
\lbrace G^{+(\I)}_{+,r} , G^{+(\II)}_{-,s} \rbrace + \lbrace  G^{+(\II)}_{+,r}, G^{+(\I)}_{-,s}  \rbrace
\label{gpgpp}
\eea
Partial spectral flowing the LHS by $\a_\I$ yields 
\bea
\text{LHS}
&\to& \lbrace G^{+(\I)}_{+,r+\a_\I/2}+ G^{+(\II)}_{+,r} , G^{+(\I)}_{-,s+\a_\I/2}+G^{+(\II)}_{-,s} \rbrace\cr
&=&
(r-s+\a_\I/2)J^+_{r+s+\a_\I/2}/2 + (r-s-\a_\I/2)J^+_{r+s+\a_\I/2}/2\cr
&=& (r-s)J^+_{r+s+\a_\I/2}
\eea
The RHS partial spectral flows to
\bea
\text{RHS}\to  (r-s)J^+_{r+s+\a_\I/2}
\eea
Thus partial spectral flow of the LHS and the RHS are equal.

Similarly the anticommutator
\be
\lbrace G^{-}_{+,r} , G^{-}_{-,s} \rbrace = - (r-s)J^-_{r+s} 
\ee
is invariant under partial spectral flow.

\b

(iv) Consider the commutator
\bea
[L_m,L_n] = {c\over12}m(m^2-1)\delta_{m+n,0}+ (m-n)L_{m+n}
\label{ll}
\eea
The LHS can be written as
\bea
[L_m,L_n] &=& [L^{(\I)}_m +L^{(\II)}_m, L^{(\I)}_n+L^{(\II)}_n]\cr
&=& [L^{(\I)}_m, L^{(\I)}_n] + [L^{(\II)}_m, L^{(\II)}_n]
\eea
Under partial spectral flow by $\a_\I$
\bea
[L^{(\I)}_m , L^{(\I)}_n]&\to& {c/2\over12}m(m^2-1)\delta_{m+n,0}+ (m-n)(L^{(\I)}_{m+n}+\a_\I J^{3(\I)}_{m+n}+\frac{\alpha_{\I}^2 c}{48}\delta_{m+n,0})\cr
[L^{(\II)}_m  ,L^{(\II)}_n]&\to& {c/2\over12}m(m^2-1)\delta_{m+n,0}+ (m-n)L^{(\II)}_{m+n}
\eea
Putting these results together, partial spectral flow of the LHS is 
\be
\text{LHS}\to{c\over12}m(m^2-1)\delta_{m+n,0}+ (m-n)(L^{(\I)}_{m+n}+\a_\I J^{3(\I)}_{m+n}+\frac{\alpha_{\I}^2 c}{48}\delta_{m+n,0}+L^{(\II)}_{m+n})
\ee
Partial spectral flow of the RHS is
\bea
\text{RHS}\to{c\over12}m(m^2-1)\delta_{m+n,0}+ (m-n)(L^{(\I)}_{m+n}+\a_\I J^{3(\I)}_{m+n}+\frac{\alpha_{\I}^2 c}{48}\delta_{m+n,0}+L^{(\II)}_{m+n})
\eea
Thus partial spectral flow of the LHS and the RHS are equal.

\bibliographystyle{JHEP}
\bibliography{bibliography.bib}

\providecommand{\href}[2]{#2}\begingroup\raggedright\begin{thebibliography}{10}

\bibitem{Maldacena:1997re}
J.M.~Maldacena, \emph{{The Large N limit of superconformal field theories and
  supergravity}}, \href{https://doi.org/10.1023/A:1026654312961}{\emph{Adv.
  Theor. Math. Phys.} {\bfseries 2} (1998) 231}
  [\href{https://arxiv.org/abs/hep-th/9711200}{{\ttfamily hep-th/9711200}}].

\bibitem{Gubser:1998bc}
S.S.~Gubser, I.R.~Klebanov and A.M.~Polyakov, \emph{{Gauge theory correlators
  from noncritical string theory}},
  \href{https://doi.org/10.1016/S0370-2693(98)00377-3}{\emph{Phys. Lett. B}
  {\bfseries 428} (1998) 105}
  [\href{https://arxiv.org/abs/hep-th/9802109}{{\ttfamily hep-th/9802109}}].

\bibitem{Witten:1998qj}
E.~Witten, \emph{{Anti-de Sitter space and holography}},
  \href{https://doi.org/10.4310/ATMP.1998.v2.n2.a2}{\emph{Adv. Theor. Math.
  Phys.} {\bfseries 2} (1998) 253}
  [\href{https://arxiv.org/abs/hep-th/9802150}{{\ttfamily hep-th/9802150}}].

\bibitem{Callan:1996dv}
C.G.~Callan and J.M.~Maldacena, \emph{{D-brane approach to black hole quantum
  mechanics}}, \href{https://doi.org/10.1016/0550-3213(96)00225-8}{\emph{Nucl.
  Phys. B} {\bfseries 472} (1996) 591}
  [\href{https://arxiv.org/abs/hep-th/9602043}{{\ttfamily hep-th/9602043}}].

\bibitem{Das:1996wn}
S.R.~Das and S.D.~Mathur, \emph{{Comparing decay rates for black holes and
  D-branes}}, \href{https://doi.org/10.1016/0550-3213(96)00453-1}{\emph{Nucl.
  Phys. B} {\bfseries 478} (1996) 561}
  [\href{https://arxiv.org/abs/hep-th/9606185}{{\ttfamily hep-th/9606185}}].

\bibitem{Das:1996ug}
S.R.~Das and S.D.~Mathur, \emph{{Excitations of D strings, entropy and
  duality}}, \href{https://doi.org/10.1016/0370-2693(96)00242-0}{\emph{Phys.
  Lett. B} {\bfseries 375} (1996) 103}
  [\href{https://arxiv.org/abs/hep-th/9601152}{{\ttfamily hep-th/9601152}}].

\bibitem{Maldacena:1996ix}
J.M.~Maldacena and A.~Strominger, \emph{{Black hole grey body factors and
  d-brane spectroscopy}},
  \href{https://doi.org/10.1103/PhysRevD.55.861}{\emph{Phys. Rev. D} {\bfseries
  55} (1997) 861} [\href{https://arxiv.org/abs/hep-th/9609026}{{\ttfamily
  hep-th/9609026}}].

\bibitem{David:1999ec}
J.R.~David, G.~Mandal and S.R.~Wadia, \emph{{D1 / D5 moduli in SCFT and gauge
  theory, and Hawking radiation}},
  \href{https://doi.org/10.1016/S0550-3213(99)00620-3}{\emph{Nucl. Phys. B}
  {\bfseries 564} (2000) 103}
  [\href{https://arxiv.org/abs/hep-th/9907075}{{\ttfamily hep-th/9907075}}].

\bibitem{Seiberg:1999xz}
N.~Seiberg and E.~Witten, \emph{{The D1 / D5 system and singular CFT}},
  \href{https://doi.org/10.1088/1126-6708/1999/04/017}{\emph{JHEP} {\bfseries
  04} (1999) 017} [\href{https://arxiv.org/abs/hep-th/9903224}{{\ttfamily
  hep-th/9903224}}].

\bibitem{Dijkgraaf:1998gf}
R.~Dijkgraaf, \emph{{Instanton strings and hyperKahler geometry}},
  \href{https://doi.org/10.1016/S0550-3213(98)00869-4}{\emph{Nucl. Phys. B}
  {\bfseries 543} (1999) 545}
  [\href{https://arxiv.org/abs/hep-th/9810210}{{\ttfamily hep-th/9810210}}].

\bibitem{Larsen:1999uk}
F.~Larsen and E.J.~Martinec, \emph{{U(1) charges and moduli in the D1 - D5
  system}}, \href{https://doi.org/10.1088/1126-6708/1999/06/019}{\emph{JHEP}
  {\bfseries 06} (1999) 019}
  [\href{https://arxiv.org/abs/hep-th/9905064}{{\ttfamily hep-th/9905064}}].

\bibitem{Jevicki:1998bm}
A.~Jevicki, M.~Mihailescu and S.~Ramgoolam, \emph{{Gravity from CFT on S**N(X):
  Symmetries and interactions}},
  \href{https://doi.org/10.1016/S0550-3213(00)00147-4}{\emph{Nucl. Phys. B}
  {\bfseries 577} (2000) 47}
  [\href{https://arxiv.org/abs/hep-th/9907144}{{\ttfamily hep-th/9907144}}].

\bibitem{Strominger:1996sh}
A.~Strominger and C.~Vafa, \emph{{Microscopic origin of the Bekenstein-Hawking
  entropy}}, \href{https://doi.org/10.1016/0370-2693(96)00345-0}{\emph{Phys.
  Lett. B} {\bfseries 379} (1996) 99}
  [\href{https://arxiv.org/abs/hep-th/9601029}{{\ttfamily hep-th/9601029}}].

\bibitem{Maldacena:1999bp}
J.M.~Maldacena, G.W.~Moore and A.~Strominger, \emph{{Counting BPS black holes
  in toroidal Type II string theory}},
  \href{https://arxiv.org/abs/hep-th/9903163}{{\ttfamily hep-th/9903163}}.

\bibitem{Lunin:2001fv}
O.~Lunin and S.D.~Mathur, \emph{{Metric of the multiply wound rotating
  string}}, \href{https://doi.org/10.1016/S0550-3213(01)00321-2}{\emph{Nucl.
  Phys. B} {\bfseries 610} (2001) 49}
  [\href{https://arxiv.org/abs/hep-th/0105136}{{\ttfamily hep-th/0105136}}].

\bibitem{Lunin:2001jy}
O.~Lunin and S.D.~Mathur, \emph{{AdS / CFT duality and the black hole
  information paradox}},
  \href{https://doi.org/10.1016/S0550-3213(01)00620-4}{\emph{Nucl. Phys. B}
  {\bfseries 623} (2002) 342}
  [\href{https://arxiv.org/abs/hep-th/0109154}{{\ttfamily hep-th/0109154}}].

\bibitem{Mathur:2005zp}
S.D.~Mathur, \emph{{The Fuzzball proposal for black holes: An Elementary
  review}}, \href{https://doi.org/10.1002/prop.200410203}{\emph{Fortsch. Phys.}
  {\bfseries 53} (2005) 793}
  [\href{https://arxiv.org/abs/hep-th/0502050}{{\ttfamily hep-th/0502050}}].

\bibitem{Bena:2015bea}
I.~Bena, S.~Giusto, R.~Russo, M.~Shigemori and N.P.~Warner, \emph{{Habemus
  Superstratum! A constructive proof of the existence of superstrata}},
  \href{https://doi.org/10.1007/JHEP05(2015)110}{\emph{JHEP} {\bfseries 05}
  (2015) 110} [\href{https://arxiv.org/abs/1503.01463}{{\ttfamily
  1503.01463}}].

\bibitem{Bena:2016agb}
I.~Bena, E.~Martinec, D.~Turton and N.P.~Warner, \emph{{Momentum Fractionation
  on Superstrata}}, \href{https://doi.org/10.1007/JHEP05(2016)064}{\emph{JHEP}
  {\bfseries 05} (2016) 064}
  [\href{https://arxiv.org/abs/1601.05805}{{\ttfamily 1601.05805}}].

\bibitem{Bena:2016ypk}
I.~Bena, S.~Giusto, E.J.~Martinec, R.~Russo, M.~Shigemori, D.~Turton et~al.,
  \emph{{Smooth horizonless geometries deep inside the black-hole regime}},
  \href{https://doi.org/10.1103/PhysRevLett.117.201601}{\emph{Phys. Rev. Lett.}
  {\bfseries 117} (2016) 201601}
  [\href{https://arxiv.org/abs/1607.03908}{{\ttfamily 1607.03908}}].

\bibitem{Bena:2017xbt}
I.~Bena, S.~Giusto, E.J.~Martinec, R.~Russo, M.~Shigemori, D.~Turton et~al.,
  \emph{{Asymptotically-flat supergravity solutions deep inside the black-hole
  regime}}, \href{https://doi.org/10.1007/JHEP02(2018)014}{\emph{JHEP}
  {\bfseries 02} (2018) 014}
  [\href{https://arxiv.org/abs/1711.10474}{{\ttfamily 1711.10474}}].

\bibitem{Ceplak:2018pws}
N.~\v{C}eplak, R.~Russo and M.~Shigemori, \emph{{Supercharging Superstrata}},
  \href{https://doi.org/10.1007/JHEP03(2019)095}{\emph{JHEP} {\bfseries 03}
  (2019) 095} [\href{https://arxiv.org/abs/1812.08761}{{\ttfamily
  1812.08761}}].

\bibitem{Heidmann:2019zws}
P.~Heidmann and N.P.~Warner, \emph{{Superstratum Symbiosis}},
  \href{https://doi.org/10.1007/JHEP09(2019)059}{\emph{JHEP} {\bfseries 09}
  (2019) 059} [\href{https://arxiv.org/abs/1903.07631}{{\ttfamily
  1903.07631}}].

\bibitem{Kanitscheider:2006zf}
I.~Kanitscheider, K.~Skenderis and M.~Taylor, \emph{{Holographic anatomy of
  fuzzballs}}, \href{https://doi.org/10.1088/1126-6708/2007/04/023}{\emph{JHEP}
  {\bfseries 04} (2007) 023}
  [\href{https://arxiv.org/abs/hep-th/0611171}{{\ttfamily hep-th/0611171}}].

\bibitem{Kanitscheider:2007wq}
I.~Kanitscheider, K.~Skenderis and M.~Taylor, \emph{{Fuzzballs with internal
  excitations}},
  \href{https://doi.org/10.1088/1126-6708/2007/06/056}{\emph{JHEP} {\bfseries
  06} (2007) 056} [\href{https://arxiv.org/abs/0704.0690}{{\ttfamily
  0704.0690}}].

\bibitem{Taylor:2007hs}
M.~Taylor, \emph{{Matching of correlators in AdS(3) / CFT(2)}},
  \href{https://doi.org/10.1088/1126-6708/2008/06/010}{\emph{JHEP} {\bfseries
  06} (2008) 010} [\href{https://arxiv.org/abs/0709.1838}{{\ttfamily
  0709.1838}}].

\bibitem{Giusto:2015dfa}
S.~Giusto, E.~Moscato and R.~Russo, \emph{{AdS$_{3}$ holography for 1/4 and 1/8
  BPS geometries}}, \href{https://doi.org/10.1007/JHEP11(2015)004}{\emph{JHEP}
  {\bfseries 11} (2015) 004}
  [\href{https://arxiv.org/abs/1507.00945}{{\ttfamily 1507.00945}}].

\bibitem{GarciaTormo:2019inl}
J.~Garcia~i Tormo and M.~Taylor, \emph{{One point functions for black hole
  microstates}}, \href{https://doi.org/10.1007/s10714-019-2566-6}{\emph{Gen.
  Rel. Grav.} {\bfseries 51} (2019) 89}
  [\href{https://arxiv.org/abs/1904.10200}{{\ttfamily 1904.10200}}].

\bibitem{Giusto:2019qig}
S.~Giusto, S.~Rawash and D.~Turton, \emph{{Ads$_{3}$ holography at dimension
  two}}, \href{https://doi.org/10.1007/JHEP07(2019)171}{\emph{JHEP} {\bfseries
  07} (2019) 171} [\href{https://arxiv.org/abs/1904.12880}{{\ttfamily
  1904.12880}}].

\bibitem{Rawash:2021pik}
S.~Rawash and D.~Turton, \emph{{Supercharged AdS$_{3}$ Holography}},
  \href{https://doi.org/10.1007/JHEP07(2021)178}{\emph{JHEP} {\bfseries 07}
  (2021) 178} [\href{https://arxiv.org/abs/2105.13046}{{\ttfamily
  2105.13046}}].

\bibitem{Ganchev:2021ewa}
B.~Ganchev, S.~Giusto, A.~Houppe and R.~Russo, \emph{{AdS$_3$ holography for
  non-BPS geometries}},  \href{https://arxiv.org/abs/2112.03287}{{\ttfamily
  2112.03287}}.

\bibitem{Eberhardt:2018ouy}
L.~Eberhardt, M.R.~Gaberdiel and R.~Gopakumar, \emph{{The Worldsheet Dual of
  the Symmetric Product CFT}},
  \href{https://doi.org/10.1007/JHEP04(2019)103}{\emph{JHEP} {\bfseries 04}
  (2019) 103} [\href{https://arxiv.org/abs/1812.01007}{{\ttfamily
  1812.01007}}].

\bibitem{Eberhardt:2019ywk}
L.~Eberhardt, M.R.~Gaberdiel and R.~Gopakumar, \emph{{Deriving the
  AdS$_{3}$/CFT$_{2}$ correspondence}},
  \href{https://doi.org/10.1007/JHEP02(2020)136}{\emph{JHEP} {\bfseries 02}
  (2020) 136} [\href{https://arxiv.org/abs/1911.00378}{{\ttfamily
  1911.00378}}].

\bibitem{Eberhardt:2020akk}
L.~Eberhardt, \emph{{AdS$_{3}$/CFT$_{2}$ at higher genus}},
  \href{https://doi.org/10.1007/JHEP05(2020)150}{\emph{JHEP} {\bfseries 05}
  (2020) 150} [\href{https://arxiv.org/abs/2002.11729}{{\ttfamily
  2002.11729}}].

\bibitem{Dei:2020zui}
A.~Dei, M.R.~Gaberdiel, R.~Gopakumar and B.~Knighton, \emph{{Free field
  world-sheet correlators for ${\rm AdS}_3$}},
  \href{https://doi.org/10.1007/JHEP02(2021)081}{\emph{JHEP} {\bfseries 02}
  (2021) 081} [\href{https://arxiv.org/abs/2009.11306}{{\ttfamily
  2009.11306}}].

\bibitem{Knighton:2020kuh}
B.~Knighton, \emph{{Higher genus correlators for tensionless AdS$_{3}$
  strings}}, \href{https://doi.org/10.1007/JHEP04(2021)211}{\emph{JHEP}
  {\bfseries 04} (2021) 211}
  [\href{https://arxiv.org/abs/2012.01445}{{\ttfamily 2012.01445}}].

\bibitem{Gaberdiel:2021kkp}
M.R.~Gaberdiel, B.~Knighton and J.~Vo\v{s}mera, \emph{{D-branes in
  $\mathrm{AdS}_3\times \mathrm{S}^3\times \mathbb{T}^4$ at $k=1$ and their
  holographic duals}},  \href{https://arxiv.org/abs/2110.05509}{{\ttfamily
  2110.05509}}.

\bibitem{Eberhardt:2021vsx}
L.~Eberhardt, \emph{{A perturbative CFT dual for pure NS-NS AdS$_3$ strings}},
  \href{https://arxiv.org/abs/2110.07535}{{\ttfamily 2110.07535}}.

\bibitem{Schwimmer:1986mf}
A.~Schwimmer and N.~Seiberg, \emph{{Comments on the N=2, N=3, N=4
  Superconformal Algebras in Two-Dimensions}},
  \href{https://doi.org/10.1016/0370-2693(87)90566-1}{\emph{Phys. Lett. B}
  {\bfseries 184} (1987) 191}.

\bibitem{Sevrin:1988ew}
A.~Sevrin, W.~Troost and A.~Van~Proeyen, \emph{{Superconformal Algebras in
  Two-Dimensions with N=4}},
  \href{https://doi.org/10.1016/0370-2693(88)90645-4}{\emph{Phys. Lett. B}
  {\bfseries 208} (1988) 447}.

\bibitem{Jejjala:2005yu}
V.~Jejjala, O.~Madden, S.F.~Ross and G.~Titchener, \emph{{Non-supersymmetric
  smooth geometries and D1-D5-P bound states}},
  \href{https://doi.org/10.1103/PhysRevD.71.124030}{\emph{Phys. Rev. D}
  {\bfseries 71} (2005) 124030}
  [\href{https://arxiv.org/abs/hep-th/0504181}{{\ttfamily hep-th/0504181}}].

\bibitem{Giusto:2012yz}
S.~Giusto, O.~Lunin, S.D.~Mathur and D.~Turton, \emph{{D1-D5-P microstates at
  the cap}}, \href{https://doi.org/10.1007/JHEP02(2013)050}{\emph{JHEP}
  {\bfseries 02} (2013) 050} [\href{https://arxiv.org/abs/1211.0306}{{\ttfamily
  1211.0306}}].

\bibitem{Chakrabarty:2015foa}
B.~Chakrabarty, D.~Turton and A.~Virmani, \emph{{Holographic description of
  non-supersymmetric orbifolded D1-D5-P solutions}},
  \href{https://doi.org/10.1007/JHEP11(2015)063}{\emph{JHEP} {\bfseries 11}
  (2015) 063} [\href{https://arxiv.org/abs/1508.01231}{{\ttfamily
  1508.01231}}].

\bibitem{Gomis:2002qi}
J.~Gomis, L.~Motl and A.~Strominger, \emph{{PP wave / CFT(2) duality}},
  \href{https://doi.org/10.1088/1126-6708/2002/11/016}{\emph{JHEP} {\bfseries
  11} (2002) 016} [\href{https://arxiv.org/abs/hep-th/0206166}{{\ttfamily
  hep-th/0206166}}].

\bibitem{Gava:2002xb}
E.~Gava and K.S.~Narain, \emph{{Proving the PP wave / CFT(2) duality}},
  \href{https://doi.org/10.1088/1126-6708/2002/12/023}{\emph{JHEP} {\bfseries
  12} (2002) 023} [\href{https://arxiv.org/abs/hep-th/0208081}{{\ttfamily
  hep-th/0208081}}].

\bibitem{Avery:2010qw}
S.G.~Avery, \emph{{Using the D1D5 CFT to Understand Black Holes}},  other
  thesis, 12, 2010, [\href{https://arxiv.org/abs/1012.0072}{{\ttfamily
  1012.0072}}].

\bibitem{Hampton:2018ygz}
S.~Hampton, S.D.~Mathur and I.G.~Zadeh, \emph{{Lifting of D1-D5-P states}},
  \href{https://doi.org/10.1007/JHEP01(2019)075}{\emph{JHEP} {\bfseries 01}
  (2019) 075} [\href{https://arxiv.org/abs/1804.10097}{{\ttfamily
  1804.10097}}].

\bibitem{Guo:2020gxm}
B.~Guo and S.D.~Mathur, \emph{{Lifting at higher levels in the D1D5 CFT}},
  \href{https://doi.org/10.1007/JHEP11(2020)145}{\emph{JHEP} {\bfseries 11}
  (2020) 145} [\href{https://arxiv.org/abs/2008.01274}{{\ttfamily
  2008.01274}}].

\bibitem{Lunin:2000yv}
O.~Lunin and S.D.~Mathur, \emph{{Correlation functions for M**N / S(N)
  orbifolds}}, \href{https://doi.org/10.1007/s002200100431}{\emph{Commun. Math.
  Phys.} {\bfseries 219} (2001) 399}
  [\href{https://arxiv.org/abs/hep-th/0006196}{{\ttfamily hep-th/0006196}}].

\bibitem{Lunin:2001pw}
O.~Lunin and S.D.~Mathur, \emph{{Three point functions for M(N) / S(N)
  orbifolds with N=4 supersymmetry}},
  \href{https://doi.org/10.1007/s002200200638}{\emph{Commun. Math. Phys.}
  {\bfseries 227} (2002) 385}
  [\href{https://arxiv.org/abs/hep-th/0103169}{{\ttfamily hep-th/0103169}}].

\bibitem{Pakman:2009zz}
A.~Pakman, L.~Rastelli and S.S.~Razamat, \emph{{Diagrams for Symmetric Product
  Orbifolds}}, \href{https://doi.org/10.1088/1126-6708/2009/10/034}{\emph{JHEP}
  {\bfseries 10} (2009) 034} [\href{https://arxiv.org/abs/0905.3448}{{\ttfamily
  0905.3448}}].

\bibitem{Pakman:2009ab}
A.~Pakman, L.~Rastelli and S.S.~Razamat, \emph{{Extremal Correlators and
  Hurwitz Numbers in Symmetric Product Orbifolds}},
  \href{https://doi.org/10.1103/PhysRevD.80.086009}{\emph{Phys. Rev. D}
  {\bfseries 80} (2009) 086009}
  [\href{https://arxiv.org/abs/0905.3451}{{\ttfamily 0905.3451}}].

\bibitem{Pakman:2009mi}
A.~Pakman, L.~Rastelli and S.S.~Razamat, \emph{{A Spin Chain for the Symmetric
  Product CFT(2)}}, \href{https://doi.org/10.1007/JHEP05(2010)099}{\emph{JHEP}
  {\bfseries 05} (2010) 099} [\href{https://arxiv.org/abs/0912.0959}{{\ttfamily
  0912.0959}}].

\bibitem{Burrington:2012yq}
B.A.~Burrington, A.W.~Peet and I.G.~Zadeh, \emph{{Operator mixing for string
  states in the D1-D5 CFT near the orbifold point}},
  \href{https://doi.org/10.1103/PhysRevD.87.106001}{\emph{Phys. Rev. D}
  {\bfseries 87} (2013) 106001}
  [\href{https://arxiv.org/abs/1211.6699}{{\ttfamily 1211.6699}}].

\bibitem{Burrington:2012yn}
B.A.~Burrington, A.W.~Peet and I.G.~Zadeh, \emph{{Twist-nontwist correlators in
  $M^N/S_N$ orbifold CFTs}},
  \href{https://doi.org/10.1103/PhysRevD.87.106008}{\emph{Phys. Rev. D}
  {\bfseries 87} (2013) 106008}
  [\href{https://arxiv.org/abs/1211.6689}{{\ttfamily 1211.6689}}].

\bibitem{Carson:2017byr}
Z.~Carson, I.T.~Jardine and A.W.~Peet, \emph{{Component twist method for higher
  twists in D1-D5 CFT}},
  \href{https://doi.org/10.1103/PhysRevD.96.026006}{\emph{Phys. Rev. D}
  {\bfseries 96} (2017) 026006}
  [\href{https://arxiv.org/abs/1704.03401}{{\ttfamily 1704.03401}}].

\bibitem{Burrington:2017jhh}
B.A.~Burrington, I.T.~Jardine and A.W.~Peet, \emph{{Operator mixing in deformed
  D1D5 CFT and the OPE on the cover}},
  \href{https://doi.org/10.1007/JHEP06(2017)149}{\emph{JHEP} {\bfseries 06}
  (2017) 149} [\href{https://arxiv.org/abs/1703.04744}{{\ttfamily
  1703.04744}}].

\bibitem{Burrington:2018upk}
B.A.~Burrington, I.T.~Jardine and A.W.~Peet, \emph{{The OPE of bare twist
  operators in bosonic $S_N$ orbifold CFTs at large $N$}},
  \href{https://doi.org/10.1007/JHEP08(2018)202}{\emph{JHEP} {\bfseries 08}
  (2018) 202} [\href{https://arxiv.org/abs/1804.01562}{{\ttfamily
  1804.01562}}].

\bibitem{Keller:2019yrr}
C.A.~Keller and I.G.~Zadeh, \emph{{Conformal Perturbation Theory for Twisted
  Fields}}, \href{https://doi.org/10.1088/1751-8121/ab6b91}{\emph{J. Phys. A}
  {\bfseries 53} (2020) 095401}
  [\href{https://arxiv.org/abs/1907.08207}{{\ttfamily 1907.08207}}].

\bibitem{Dei:2019iym}
A.~Dei and L.~Eberhardt, \emph{{Correlators of the symmetric product
  orbifold}}, \href{https://doi.org/10.1007/JHEP01(2020)108}{\emph{JHEP}
  {\bfseries 01} (2020) 108}
  [\href{https://arxiv.org/abs/1911.08485}{{\ttfamily 1911.08485}}].

\bibitem{Avery:2010er}
S.G.~Avery, B.D.~Chowdhury and S.D.~Mathur, \emph{{Deforming the D1D5 CFT away
  from the orbifold point}},
  \href{https://doi.org/10.1007/JHEP06(2010)031}{\emph{JHEP} {\bfseries 06}
  (2010) 031} [\href{https://arxiv.org/abs/1002.3132}{{\ttfamily 1002.3132}}].

\bibitem{Avery:2010hs}
S.G.~Avery, B.D.~Chowdhury and S.D.~Mathur, \emph{{Excitations in the deformed
  D1D5 CFT}}, \href{https://doi.org/10.1007/JHEP06(2010)032}{\emph{JHEP}
  {\bfseries 06} (2010) 032} [\href{https://arxiv.org/abs/1003.2746}{{\ttfamily
  1003.2746}}].

\bibitem{Burrington:2014yia}
B.A.~Burrington, S.D.~Mathur, A.W.~Peet and I.G.~Zadeh, \emph{{Analyzing the
  squeezed state generated by a twist deformation}},
  \href{https://doi.org/10.1103/PhysRevD.91.124072}{\emph{Phys. Rev. D}
  {\bfseries 91} (2015) 124072}
  [\href{https://arxiv.org/abs/1410.5790}{{\ttfamily 1410.5790}}].

\bibitem{Carson:2014yxa}
Z.~Carson, S.~Hampton, S.D.~Mathur and D.~Turton, \emph{{Effect of the twist
  operator in the D1D5 CFT}},
  \href{https://doi.org/10.1007/JHEP08(2014)064}{\emph{JHEP} {\bfseries 08}
  (2014) 064} [\href{https://arxiv.org/abs/1405.0259}{{\ttfamily 1405.0259}}].

\bibitem{Carson:2014xwa}
Z.~Carson, S.D.~Mathur and D.~Turton, \emph{{Bogoliubov coefficients for the
  twist operator in the D1D5 CFT}},
  \href{https://doi.org/10.1016/j.nuclphysb.2014.10.018}{\emph{Nucl. Phys. B}
  {\bfseries 889} (2014) 443}
  [\href{https://arxiv.org/abs/1406.6977}{{\ttfamily 1406.6977}}].

\bibitem{Carson:2014ena}
Z.~Carson, S.~Hampton, S.D.~Mathur and D.~Turton, \emph{{Effect of the
  deformation operator in the D1D5 CFT}},
  \href{https://doi.org/10.1007/JHEP01(2015)071}{\emph{JHEP} {\bfseries 01}
  (2015) 071} [\href{https://arxiv.org/abs/1410.4543}{{\ttfamily 1410.4543}}].

\bibitem{Carson:2016cjj}
Z.~Carson, S.~Hampton and S.D.~Mathur, \emph{{One-Loop Transition Amplitudes in
  the D1D5 CFT}}, \href{https://doi.org/10.1007/JHEP01(2017)006}{\emph{JHEP}
  {\bfseries 01} (2017) 006}
  [\href{https://arxiv.org/abs/1606.06212}{{\ttfamily 1606.06212}}].

\bibitem{Carson:2016uwf}
Z.~Carson, S.~Hampton and S.D.~Mathur, \emph{{Full action of two deformation
  operators in the D1D5 CFT}},
  \href{https://doi.org/10.1007/JHEP11(2017)096}{\emph{JHEP} {\bfseries 11}
  (2017) 096} [\href{https://arxiv.org/abs/1612.03886}{{\ttfamily
  1612.03886}}].

\bibitem{Carson:2015ohj}
Z.~Carson, S.~Hampton and S.D.~Mathur, \emph{{Second order effect of twist
  deformations in the D1D5 CFT}},
  \href{https://doi.org/10.1007/JHEP04(2016)115}{\emph{JHEP} {\bfseries 04}
  (2016) 115} [\href{https://arxiv.org/abs/1511.04046}{{\ttfamily
  1511.04046}}].

\bibitem{Hampton:2019csz}
S.~Hampton and S.D.~Mathur, \emph{{Thermalization in the D1D5 CFT}},
  \href{https://doi.org/10.1007/JHEP06(2020)004}{\emph{JHEP} {\bfseries 06}
  (2020) 004} [\href{https://arxiv.org/abs/1910.01690}{{\ttfamily
  1910.01690}}].

\bibitem{Guo:2021ybz}
B.~Guo and S.~Hampton, \emph{{A freely falling graviton in the D1D5 CFT}},
  \href{https://arxiv.org/abs/2107.11883}{{\ttfamily 2107.11883}}.

\bibitem{Guo:2021gqd}
B.~Guo and S.~Hampton, \emph{{The Dual of a Tidal Force in the D1D5 CFT}},
  \href{https://arxiv.org/abs/2108.00068}{{\ttfamily 2108.00068}}.

\bibitem{Guo:2019pzk}
B.~Guo and S.D.~Mathur, \emph{{Lifting of states in 2-dimensional $N = 4$
  supersymmetric CFTs}},
  \href{https://doi.org/10.1007/JHEP10(2019)155}{\emph{JHEP} {\bfseries 10}
  (2019) 155} [\href{https://arxiv.org/abs/1905.11923}{{\ttfamily
  1905.11923}}].

\bibitem{Gaberdiel:2015uca}
M.R.~Gaberdiel, C.~Peng and I.G.~Zadeh, \emph{{Higgsing the stringy higher spin
  symmetry}}, \href{https://doi.org/10.1007/JHEP10(2015)101}{\emph{JHEP}
  {\bfseries 10} (2015) 101}
  [\href{https://arxiv.org/abs/1506.02045}{{\ttfamily 1506.02045}}].

\bibitem{Guo:2019ady}
B.~Guo and S.D.~Mathur, \emph{{Lifting of level-1 states in the D1D5 CFT}},
  \href{https://doi.org/10.1007/JHEP03(2020)028}{\emph{JHEP} {\bfseries 03}
  (2020) 028} [\href{https://arxiv.org/abs/1912.05567}{{\ttfamily
  1912.05567}}].

\bibitem{Lima:2020boh}
A.A.~Lima, G.M.~Sotkov and M.~Stanishkov, \emph{{Microstate Renormalization in
  Deformed D1-D5 SCFT}},
  \href{https://doi.org/10.1016/j.physletb.2020.135630}{\emph{Phys. Lett. B}
  {\bfseries 808} (2020) 135630}
  [\href{https://arxiv.org/abs/2005.06702}{{\ttfamily 2005.06702}}].

\bibitem{Lima:2020kek}
A.A.~Lima, G.M.~Sotkov and M.~Stanishkov, \emph{{Renormalization of twisted
  Ramond fields in D1-D5 SCFT$_{2}$}},
  \href{https://doi.org/10.1007/JHEP03(2021)202}{\emph{JHEP} {\bfseries 03}
  (2021) 202} [\href{https://arxiv.org/abs/2010.00172}{{\ttfamily
  2010.00172}}].

\bibitem{Lima:2020nnx}
A.A.~Lima, G.M.~Sotkov and M.~Stanishkov, \emph{{Correlation functions of
  composite Ramond fields in deformed D1-D5 orbifold SCFT$_2$}},
  \href{https://doi.org/10.1103/PhysRevD.102.106004}{\emph{Phys. Rev. D}
  {\bfseries 102} (2020) 106004}
  [\href{https://arxiv.org/abs/2006.16303}{{\ttfamily 2006.16303}}].

\bibitem{Lima:2020urq}
A.A.~Lima, G.M.~Sotkov and M.~Stanishkov, \emph{{Dynamics of R-neutral Ramond
  fields in the D1-D5 SCFT}},
  \href{https://arxiv.org/abs/2012.08021}{{\ttfamily 2012.08021}}.

\bibitem{Lima:2021wrz}
A.A.~Lima, G.M.~Sotkov and M.~Stanishkov, \emph{{On the Dynamics of Protected
  Ramond Ground States in the D1-D5 CFT}},
  \href{https://arxiv.org/abs/2103.04459}{{\ttfamily 2103.04459}}.

\bibitem{Benjamin:2021zkn}
N.~Benjamin, C.A.~Keller and I.G.~Zadeh, \emph{{Lifting 1/4-BPS states in
  $AdS_{3}\times S^{3}\times T^{4}$}},
  \href{https://doi.org/10.1007/JHEP10(2021)089}{\emph{JHEP} {\bfseries 10}
  (2021) 089} [\href{https://arxiv.org/abs/2107.00655}{{\ttfamily
  2107.00655}}].

\bibitem{Prokushkin:1998bq}
S.F.~Prokushkin and M.A.~Vasiliev, \emph{{Higher spin gauge interactions for
  massive matter fields in 3-D AdS space-time}},
  \href{https://doi.org/10.1016/S0550-3213(98)00839-6}{\emph{Nucl. Phys. B}
  {\bfseries 545} (1999) 385}
  [\href{https://arxiv.org/abs/hep-th/9806236}{{\ttfamily hep-th/9806236}}].

\bibitem{Vasiliev:2003ev}
M.A.~Vasiliev, \emph{{Nonlinear equations for symmetric massless higher spin
  fields in (A)dS(d)}},
  \href{https://doi.org/10.1016/S0370-2693(03)00872-4}{\emph{Phys. Lett. B}
  {\bfseries 567} (2003) 139}
  [\href{https://arxiv.org/abs/hep-th/0304049}{{\ttfamily hep-th/0304049}}].

\bibitem{Gaberdiel:2014cha}
M.R.~Gaberdiel and R.~Gopakumar, \emph{{Higher Spins \& Strings}},
  \href{https://doi.org/10.1007/JHEP11(2014)044}{\emph{JHEP} {\bfseries 11}
  (2014) 044} [\href{https://arxiv.org/abs/1406.6103}{{\ttfamily 1406.6103}}].

\bibitem{Gaberdiel:2015mra}
M.R.~Gaberdiel and R.~Gopakumar, \emph{{Stringy Symmetries and the Higher Spin
  Square}}, \href{https://doi.org/10.1088/1751-8113/48/18/185402}{\emph{J.
  Phys. A} {\bfseries 48} (2015) 185402}
  [\href{https://arxiv.org/abs/1501.07236}{{\ttfamily 1501.07236}}].

\end{thebibliography}\endgroup

\end{document}